\documentclass{aa} %
\usepackage{graphicx}
\usepackage{lscape}
\usepackage{longtable}
\usepackage{epstopdf}
\usepackage{natbib}
\usepackage{txfonts}
\begin{document}
   \title{Spectroscopy of Very Low Mass Stars and Brown Dwarfs in the Lambda Orionis Star Forming Region.}

   \subtitle{I. Enlarging the census down to the planetary mass domain in Collinder 69}

   \author{A. Bayo
          \inst{1,3}
          \and
          D. Barrado
          \inst{2,3}
          \and
          J. Stauffer
          \inst{4}
          \and          
          M. Morales-Calder\'on          
          \inst{3,4}
          \and
          C. Melo
          \inst{1}
          \and
          N. Hu\'elamo
          \inst{3}
          \and
          H. Bouy
          \inst{3}
          \and
          B. Stelzer
          \inst{5}
          \and
          M. Tamura
          \inst{6}
          \and
          R. Jayawardhana
          \inst{7}
                   }

   \institute{European Southern Observatory, Alonso de C\'ordova 3107, Vitacura, Santiago, Chile\\
              \email{abayo@eso.org}
         \and 
             Calar Alto Observatory, Centro Astron\'omico Hispano Alem\'an, C/Jes\'us Durb\'an Rem\'on, 2-2 04004 Almer\'ia, Spain
         \and    
             Depto. Astrof\'isica, Centro de Astrobiolog\'ia (INTA-CSIC), P. O. Box 78, E-28691 Villanueva de la Ca\~nada, Spain
         \and
             Spitzer Science Center, California Institute of Technology, Pasadena, CA 91125
         \and
             INAF - Osservatorio Astronomico di Palermo, Piazza del Parlamento 1, 90134 Palermo, Italy
         \and
             National Astronomical Observatory, 2-21-1 Osawa, Mitaka, Tokyo 181-8588
         \and
         	    Department of Astronomy and Astrophysics, University of Toronto, Toronto, ON M5S 3H8, Canada.}
	    
   \date{}

  \abstract
   {Whilst there is a generally accepted evolutionary scheme for the formation of low--mass stars, the analogous processes when moving down in mass to the brown dwarfs regime are not yet well understood.
   }
   {In this first paper we try to build the most complete and unbiased spectroscopically confirmed census of the population of Collinder 69, the central cluster of the Lambda Orionis star forming region, as a first step in addressing the question of how brown dwarfs and planetary mass objects form.}
   {We have studied age dependent features in optical and near-infrared spectra of candidate members to the cluster (such as alkali lines and accretion associated indicators). In addition, we have complemented that study with the analysis of other youth indicators like X-ray emission or mid-infrared excess.}
   {We have confirmed the membership to Collinder 69 of $\sim$90 photometric candidate members. As a byproduct we have determined a temperature scale for young M, very low--mass stars and brown dwarfs. We have assembled one of the most complete Initial Mass Functions from 0.016 to 20 M$_{\odot}$. And, finally, we have studied the implications of the spatial distribution of the confirmed members on the proposed mechanisms of brown dwarfs formation.}
   {}


   \maketitle
%

\section{Introduction}
\label{sec:intro}

In the current paradigm, stars are born within molecular clouds, which are accumulations of gas and dust. These clouds are initially supported against gravitational collapse by a combination of thermal, magnetic, and turbulent pressure (\citealt{Shu87}; \citealt{Mouschovias91}). Nevertheless, molecular clouds can fragment in smaller and denser cores, where the presence of gravitational instabilities yields a collapse of the cloud material (\citealt{Shu87}, and references therein).

The discovery of the first brown dwarfs in 1995 (very low mass objects characterized by the lack of stable hydrogen burning in their interior, masses typically below 0.072M$_\odot$) led to a debate on the formation mechanism of this type of objects that is still open today. Since the typical thermal Jeans mass in molecular cloud cores is around 1 M$_{\odot}$, a thermally supported cloud does not fragment in cores of substellar masses and the formation of brown dwarfs cannot be directly explained as a scaled-down version of low mass star formation.

While \citet{Padoan02} argued that brown dwarfs form via ``turbulent fragmentation'' (the density enhancements produced by the turbulence decrease the Jeans mass); \citet{Reipurth01} suggested that they may be stellar embryos ejected from newborn multiple systems before they accreted enough mass to start hydrogen burning (in this model, the truncation of the disks is explained by dynamical interactions). Besides, more recently, \cite{Whitworth04} proposed photo-evaporation of massive pre-stellar cores as the formation mechanism of brown dwarfs.

Even though a significant number of brown dwarfs with ages around a few Myr have been reported to harbour active disks \citep{Luhman97, Fernandez01, Natta04, Barrado03, Barrado04a, Mohanty05} favouring the ``in-situ" formation scenario, more homogeneous and systematic studies of disk accretion, rotation and activity (and the relationship between classic indicators of activity such as H$\alpha$ and these phenomena) in young brown dwarfs could help to confirm the ``universality" of this mechanism or the dependence on other (environmental) factors.

These kinds of studies can only be carried out setting their bases on robust census (well below the hydrogen burning limit) of star forming regions of different ages and environments. This work presents such a census for a very interesting star forming region, and, by studying the spatial distribution of its members, addresses the question on the mechanism of formation of brown dwarfs on this particular environment. 

The Lambda Orionis Star Forming Region (LOSFR) is associated to the O8III star $\lambda$ Orionis (located at $\sim$400 pc, \citealt{Murdin77}), the head of the Orion giant. It comprises both recently formed stars up to $\sim$24 M$_{\odot}$ and dark clouds actively forming stars. Although its properties (morphology, distance, reddening, size) make this star forming an ideal laboratory to test star formation theories, until recently, it had not been very well studied. 

At the beginning of the 80s, \citet{Duerr82} carried out an H$\alpha$ emission survey identifying three stellar clusters centered around the dark clouds Barnard 30 and Barnard 35. 
 Those clusters were later confirmed from a statistical point of view by \citet{Gomez98}. In particular, Collinder 69, the central one, is a well-defined, compact open cluster affected by rather low extinction A$_{\rm v}$$\sim$0.36 mag, \citealt{Duerr82}). It is quite rich, with one O binary star ($\lambda$ Ori itself), about a dozen B stars \citep{Duerr82} and a well-populated sequence of low-mass stars and brown dwarfs \citep{Barrado04b,Barrado07}. 

More recently a number of photometric studies have been published focused on Collinder 69: \citet{DM99, DM01, Dolan02} obtained optical photometry and presented a selection of candidates with estimated masses down to $\sim$0.3 M$_{\odot}$ (assuming an age of 5 Myr). \cite{MoralesPhD} presented a very complete compilation of photometry from the optical to the mid-IR obtained with different ground-based and space observatories (including previous candidates by \citealt{Barrado04, Barrado07}). They propose a list of candidates with estimated masses well below the hydrogen burning boundary if membership to the cluster is confirmed. This sample (together with the X-ray candidates, see next section) represents the starting point of this work. Our aim here is the spectroscopic characterization and confirmation of membership. When we combine our membership results with those of other spectroscopic works \citep{DM99, DM01, Barrado04, Sacco08, Maxted08} we find a pollution rate of only $\sim$9\%; this suggests that the methods followed by \cite{MoralesPhD} to obtain candidate members are very reliable.

This paper is organized as follows: In Section~\ref{sec:phot} we describe very briefly the photometric and X-ray data from which the candidate selection was obtained by \cite{MoralesPhD} and \cite{Barrado11}. In Section~\ref{sec:data} we describe the spectroscopic observations of the candidates. 
In Section~\ref{sec:teffscale} we present a temperature scale derived for M young very low mass stars and brown dwarfs (from our spectroscopically confirmed members of Collinder 69). In Section~\ref{sec:membership} we describe the procedure followed to build our final census of members. 
Finally in Sections~\ref{sec:spatialdist} and~\ref{sec:IMF} we present  the study of the spatial distribution of the confirmed members (and its implications in the theories of formation of brown dwarfs) and one of the most complete Initial Mass Functions constructed so far, respectively. Our conclusions are summarized in Section~\ref{sec:conclusions}.

On a forthcoming paper (Bayo et al. 2011, in prep., from now on Paper II), we will present properties of individual members of Collinder 69 (such as accretion rates, rotational velocities, etc.) as well as more general cluster related ones (such as disk ratios and disks spatial distribution).


\section{Photometric candidates in Collinder 69}
\label{sec:phot}

\begin{figure}[htbp]
\begin{center}
\includegraphics[width=9.4cm]{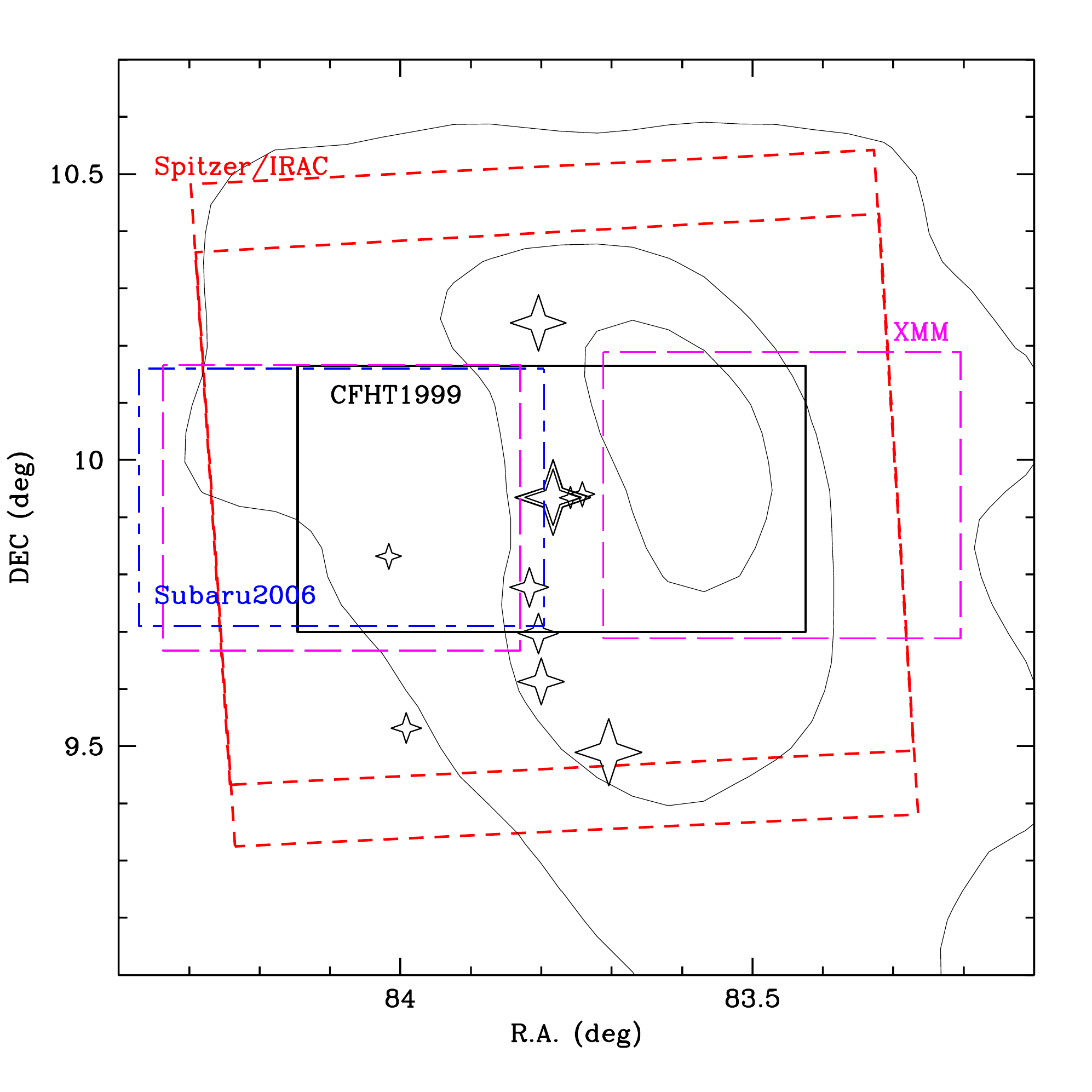}
\caption{Sketch of the areas covered by the main optical, infrared, and x-rays surveys in the region of Collinder 69. $\lambda$ Ori itself is highlighted with a large star and the B population with smaller ones. Omega 2000 and Ingrid data together fill the area covered by the CFHT survey (we do not include the mosaics for the sake of clarity in the figure). The large red dashed boxes show the IRAC FoV --the the edges are the regions with coverage only in 3.6/5.8 or 4.5/8.0-- and the area surveyed by MIPS is slightly larger.}
\label{fig:FoVs}
\end{center}
\end{figure}

As mentioned before, the base of this work is the follow-up of the photometric candidates proposed in \cite{MoralesPhD} and a sample of the X-ray emitters reported in \cite{Barrado11}. Here we provide a short description of the data used in those two works and in Fig~\ref{fig:FoVs} we show a diagram with the coverage of the different surveys over the contours of the 100$\mu$m IRAS image of the cluster: 

\begin{itemize}

\item{{\bf Optical:}

\begin{enumerate}

\item The CFHT1999 Survey (Cousins R and I bands): described in \citet{Barrado04}, for cluster members, the faint limit is set by R$_{\rm Complete}$$\sim$22.75~mag at (R - I) = 2.5, corresponding then to I$_{\rm Complete,cluster}$$\sim$20.2 mag. For a DUSTY 5 Myr isochrone (Chabrier et al. 2000), and the distance and standard extinction for the cluster, this limit corresponds to 20 M$_{Jupiter}$.

\item The Subaru2006 Survey (i' and z' bands): described in \citet{MoralesPhD}, for a DUSTY 5Myr isochrone \citep{Chabrier00}, and the distance and standard extinction for the cluster, an I magnitude of 25.5~mag corresponds to $\sim$7~M$_{Jupiter}$; for a COND 5Myr isochrone, more appropriate for this range of temperatures, the completeness limit is located at 3~M$_{Jupiter}$.
\end{enumerate}
}

\item{{\bf Near Infrared: }

\begin{enumerate}

\item 2MASS (J, H, Ks bands) provides near infrared data down to a limiting magnitude of $J$=16.8, $H$=16.1, and $Ks$=15.3 mag ($\sim$30~M$_{Jupiter}$ for a DUSTY, \citealt{Chabrier00}, 5Myr isochrone). 

\item INGRID Survey (J, H, Ks bands): described in \citet{Barrado07}, the detection limit can be estimated as J$_{Limit}$=21.1 mag and the completeness limit as J$_{Complete}$=19.5 mag. For a DUSTY 5 Myr isochrone \citep{Chabrier00}, and the distance and standard extinction for the cluster, this limit corresponds to 10~M$_{Jupiter}$.

\item Omega2000 Survey (J, H, Ks bands): described in \citet{Barrado07}, the completeness limits for the survey are: J$_{Complete}$=20 mag, H$_{Complete}$=19, and Ks$_{Complete}$=18. For a DUSTY 5Myr isochrone \citep{Chabrier00}, and the distance and standard extinction for the cluster, this limit corresponds to 8~M$_{Jupiter}$.
\end{enumerate}
}

\item{{\bf Mid Infrared:}

\begin{enumerate}
\item Spitzer mid-IR imaging (3.6, 4.5, 5.8, 8.0 and 24.0~$\mu$m): described in \citet{Barrado07} and \citet{MoralesPhD}, the completeness limits are at [3.6]$_{\rm Complete}$=16.5 mag, [4.5]$_{\rm Complete}$=16.5, [5.8]$_{\rm Complete}$=14.5, and [8.0]$_{\rm Complete}$=13.75.
For cluster members, the completeness limit at 3.6~$\mu$m (for a 5Myr isochrone by \citealt{Baraffe98}) corresponds to a mass $\sim$0.04 M$_{\odot}$.
\end{enumerate}
}
\item{{\bf X-rays:}

\begin{enumerate}
\item {\em XMM-Newton} observations of Collinder 69: The observations consisted of two fields, one to the East and another to the West of the bright star $\lambda$ Ori and allowed us to study the Weak-line TTauri population of this open cluster. A list of detections was compiled, cross-matched with our previous photometric surveys  and several selection criteria were applied to derive a catalog of new candidate members (see details of the whole process in \citealt{Barrado11}). This study produced a list of 66 candidates (19 new) from which 44 sources have been spectroscopically confirmed already by \citet{DM99, Barrado04,Sacco08, Maxted08} or this work (4 of them have only been confirmed by us). The X-ray detected cluster sample is complete down to ~$\sim$0.3M$_{\sun}$, with some detections for confirmed members with masses close to 0.1M$_{\sun}$.
\end{enumerate}
}
\end{itemize}

\section{Spectroscopic observations and data analysis}
\label{sec:data}

Regarding the spectral analysis we have used our own data (see Sections~\ref{sec:os} and~\ref{sec:nirs}) and the measurements from \citet{DM01, Barrado04, Sacco08, Maxted08}. 

In brief: \cite{DM01} obtained R$\sim$20000 spectra with a setup that allowed them to study the Li I $\lambda$6707 \AA~ and H$\alpha$ lines, and a rich array of metal lines near 6450 \AA~for precise radial velocity measurements to confirm members down to $\sim$0.3 M$_{\odot}$. Strong H$\alpha$ emission (indicative of accretion) has also been detected from three candidate substellar members in the very deep survey for young stars of \citet{Barrado04}.  Finally, \citet{Sacco08} and \citet{Maxted08} published their results on the comparative analysis of high resolution spectra of two samples of low-mass young stars (candidate members to $\sigma$ Ori and Collinder 69 clusters); with a resolution of R$\sim$16000-17000 they studied binarity, Li absorption, H$\alpha$ emission and derived rotational velocities for candidates from \citet{DM99} and \citet{Barrado04} close to the brown dwarf domain.

Because of the large amount of data we were dealing with (in particular remarkable for own our data-sets, see Tables~\ref{C69campains} and~\ref{C69_IRcampains} and more details in \citealt{BayoPhD}), we developed an automatic procedure to perform the line characterization. We also studied the effect of the spectral resolution in the measurements provided by this automatic procedure (see Appendix A).

\subsection{Optical spectroscopy}
\label{sec:os}

During the last seven years, our group has been granted time at different observatories to perform spectroscopic observations of the previously described candidates. These observations comprise a wide range of resolutions and wavelength coverages. In Table~\ref{C69campains}, the most relevant information of the different runs is displayed. 

Except where otherwise stated, we reduced all data within the IRAF\footnote{IRAF is distributed by the National Optical Astronomy Observatory, which is operated by the Association of Universities for Research in Astronomy, Inc. under contract to the National Science Foundation.} environment in standard way.

{\bf LRIS, Low Resolution Imaging Spectrometer (Keck):} We used the 1200 lines/mm grating, with a scale of  0.63 "/pixel and a resolution of $\sim$2 \AA~(measured in a NeHe comparison lamp, R$\sim$2650). The one arsec slit was used and a wavelength coverage of 6425--7692 \AA~was achieved.
During the same run we collected low-resolution spectra with the 400 l/mm grating and also the one arcsec slit. 
The wavelength calibration is better than 0.4 \AA, the resolution is 6.0 \AA~around the wavelength of H$\alpha$ as measured with a NeAr lamp (R$\sim$950), and the wavelength coverage: 6250--9600 \AA. 

{\bf MIKE, Magellan Inamori Kyocera Echelle (Magellan):} We used the prior to 2004 Red-MIKE standard configuration: R2 echelle grating, scale of $\sim$0.29"/pixe and 1.0" slit. With our set-up, the spectral coverage in the red arm was 4430--7250 \AA. In order to improve the final S/N, we degraded the resolution by rebinning the original data during the readout to 2 and 8 pixels in the spatial and spectral directions, respectively, achieving a  resolution of 0.55 \AA~(R$\sim$11250). 

{\bf B\&C, Boller \& Chivens spectrograph (Magellan):} We used the 1200 l/mm grating and the 1.0" slit; with a scale of  0.79"/pixel a resolution of $\sim$2.5 \AA~as measured in a HeNeAr comparison lamp (R$\sim$2600); and a wavelength coverage from 6200 to 7825 \AA. We also used the 300 l/mm grating (same slit as before,a resolution of $\sim$800), and a wavelength coverage 5000--10200\AA; this way, the Magellan spectra have slightly worse resolution and larger spectral range. 

{\bf FLAMES, Fibre Large Array Multi Element Spectrograph (VLT):} Program 080.C-0592. We used the LR6 grating with a measured resolution of 0.76 \AA; each fiber has an aperture of 1.2 " on the sky and GIRAFFE has a scale of 0.3"/pixel in MEDUSA (the resolution is $\sim$8600 and the wavelength range 6438--10350\AA). The data reduction was performed using the GIRAFFE gir--BLDRS pipeline vers. 1.12, following the standard steps which include 
correction for the differences in the fiber transmission.
The spectra processed by the pipeline are not corrected for sky background; therefore, the analysis that we present was performed after subtracting a sky spectrum. This background spectrum was computed as the median of those obtained from the fibers positioned "on sky". Since the nebular emission of the region of Collinder 69 is not negligible and our sky fibers were distributed quite homogeneously, we studied the variations of the H$\alpha$ nebular emission with these sky fibers to have an idea of the accuracy of the correction achieved. We measured a mean full width at 10\% of $\sim$41 km/s with a standard deviation of $\sim$3 km/s. Therefore the dispersion measured in different fibers translated into an added $\sim$7\% uncertainty in our measurements.

{\bf CAFOS, Calar Alto Faint Object Spectrograph (CAHA 2.2m telescope):} We used the R-200 grism, with scale of 0.53"/pixel and a measured resolution of $\sim$11 \AA~(we used the 1.6" slit, R$\sim$600, with a wavelength coverage of 6200--10350\AA). 

{\bf TWIN (CAHA 3.5m telescope):}  In every run we used the T-13 grating and the 1.2" and 1.5" slits and only the data coming from the red arm were processed. The pixel scale is 0.56"/pixel, and we measured a resolution of $\sim$6 \AA~(R$\sim$1100 and wavelength coverage from 5600 to 10400 \AA). 

\begin{table*}
\caption[A summary of the Collinder 69 optical spectroscopic campaigns.]{A summary of the Collinder 69 optical spectroscopic campaigns.\label{C69campains}} 
\tiny
\begin{tabular}{cccccc}
\hline\hline
Date & Observatory/Telescope/Instrument & Resolution & Wavelength &Number of sources  &Original photometric\\
          &                                                                & $\Delta\lambda/\lambda$           &   coverage & observed                   &survey              \\
\hline
Nov 2-5, 2002   & Mauna Kea / Keck / LRIS             & $\sim$2650  & 6425--7692\AA & 12         &CFHT1999\\
Nov 2-5, 2002   & Mauna Kea / Keck / LRIS             & $\sim$ 950  & 6250--9600\AA & 29         &CFHT1999\\
Dec 11-14, 2002 & Las Campanas / Magellan / MIKE  & $\sim$11250 & 4430--7250\AA & 14         &CFHT1999\\
Mar 9-11, 2003     & Las Campanas / Magellan / B\&C     & $\sim$2600  & 6200--7825\AA &  2         &CFHT1999\\
Mar 9-11, 2003     & Las Campanas / Magellan / B\&C     & $\sim$800   & 5000--10200\AA & 3         &CFHT1999\\
Nov 22-25, 2005 & CAHA / 3.5m / TWIN                    & $\sim$1100  & 5600--10425\AA & 5         &CFHT1999\\ 
Nov 20-23, 2006 & CAHA / 3.5m / TWIN                    & $\sim$1100  & 5700--9900\AA  & 8         &CFHT1999 \& $1^{o} \times 1^{o}$ Spitzer\\ 
Nov. 30 - Dec. 11, 2007     & CAHA /2.2m / CAFOS             & $\sim$600  &  6200--10350\AA& 37        &CFHT1999 \& $1^{o} \times 1^{o}$ Spitzer \& \\
& & & & & {\it XMM-Newton} survey\\ 
Jan 5, 2008     & Paranal / VLT /FLAMES                 & $\sim$8600   & 6438--7184\AA   & 40        &CFHT1999 \& $1^{o} \times 1^{o}$ Spitzer\\ 
\hline\hline
\end{tabular}
\end{table*}
\normalsize

\subsection{Near-infrared spectroscopy}
\label{sec:nirs}
{\bf SOFI, Son of ISAAC (NTT):} Program 078.C-0124. We used two low-resolution grisms with the 0.6" slit to roughly cover the $JHK$ bands on a Hawaii HgCdTe 1024$\times$1024 detector with a plate scale of 0.292$''$/pix. The blue grism covers 0.95--1.63 $\mu$m and the red grism the region between 1.53--2.52 $\mu$m. The corresponding spectral resolutions were 930 and 980 respectively. 
The telescope was nodded along the slit between two positions following the usual ABBA pattern. 

In addition to the science targets we observed several ``telluric standards" (A0V objects at similar airmasses) to remove telluric water absorption bands as described by \citet{Vacca03}, and to estimate the instrumental response.
A Xe arc lamp was used for the wavelength calibration (consistent with the OH airglow calibration) with an accuracy of 1.2 \AA{} for the blue grism, and 2 \AA{} for the red one.

{\bf NIRSPEC, the Near InfRared echelle SPECtrograph (Keck):} In both campaigns we used Nirspec-3 with the 0.57" slit, covering the 1.143-1.375 $\mu$m wavelength region. The corresponding spectral resolution is $\sim$2000 (and the pixel scale 0.18"/pixel). Again, classical ABBA pattern was followed and telluric A0V standards were observed. In this case, the data reduction process was carried out using the IDL based software REDSPEC (which performs the standard steps in an pseudo-automatic manner where some interaction with the user is required). 

{\bf IRCS, the InfraRed Camera and Spectrograph (Subaru):} We used the ``Grism HK'' covering a wavelength range of 1.4--2.5 $\mu$m at a very low resolution of $\sim$150 (0.3" slit and the 52 mas scale). The spectra of the science targets, spectral templates and A0V standards were obtained following an ABBA pattern and the reduction of the data was performed using IRAF in a standard manner. 

\begin{table*}
\caption[A summary of the Collinder 69 near infrared spectroscopic campaigns.]{A summary of the Collinder 69 near infrared spectroscopic campaigns.} \label{C69_IRcampains}
\begin{tabular}{cccccc}
\hline\hline
Date & Observatory/Telescope/Instrument & Resolution & Wavelength & Number of sources &Original photometric\\
     &                                  &    $\Delta\lambda/\lambda$         &      coverage               & observed                   &survey              \\
\hline
Dec 22-23, 2004 & Mauna Kea / Keck / NIRSPEC         & $\sim$2000  & 1.143--1.375$\mu$m & 4$^*$     &CFHT1999\\
Dec 9, 2005     & Mauna Kea / Keck / NIRSPEC         & $\sim$2000  & 1.143--1.375$\mu$m & 9$^*$     &CFHT1999\\
Jan 9-11, 2007   & La Silla / NTT /SOFI                 & $\sim$950   & 0.950--2.500$\mu$m & 2         &CFHT1999\\
Nov 10, 2008    & Mauna Kea / Subaru /IRCS              & $\sim$150   & 1.400--2.500$\mu$m & 8$^*$     &Subaru2006\\
\hline\hline
\end{tabular}

\begin{tiny}
\vspace{0.2cm}
$^*$ For some of the sources the obtained S/N ratio was too low due to poor weather conditions to perform the analysis and therefore are not listed on the tables with the results from these analyses.
\end{tiny}
\end{table*}
\normalsize

\section{Temperature Scale for young M very low-mass stars and brown dwarfs}
\label{sec:teffscale}

\subsection{Spectral Typing}
\label{sec:SpT}
Depending on the expected nature of the sources themselves and the characteristics of the available spectra, we have used different approaches to derive spectral types for our candidates: 

{\bf Optical spectra:} We considered two groups according to the effective temperature derived from the SED fit (see Sect.~\ref{sec:SED}): ``warm" (T$_{\rm eff} \gtrsim$ 4000 K, about M0 spectral type) and ``cool" (T$_{\rm eff} \lesssim$ 4000 K) sources. 

\begin{figure}
\includegraphics[width=4.46cm, clip]{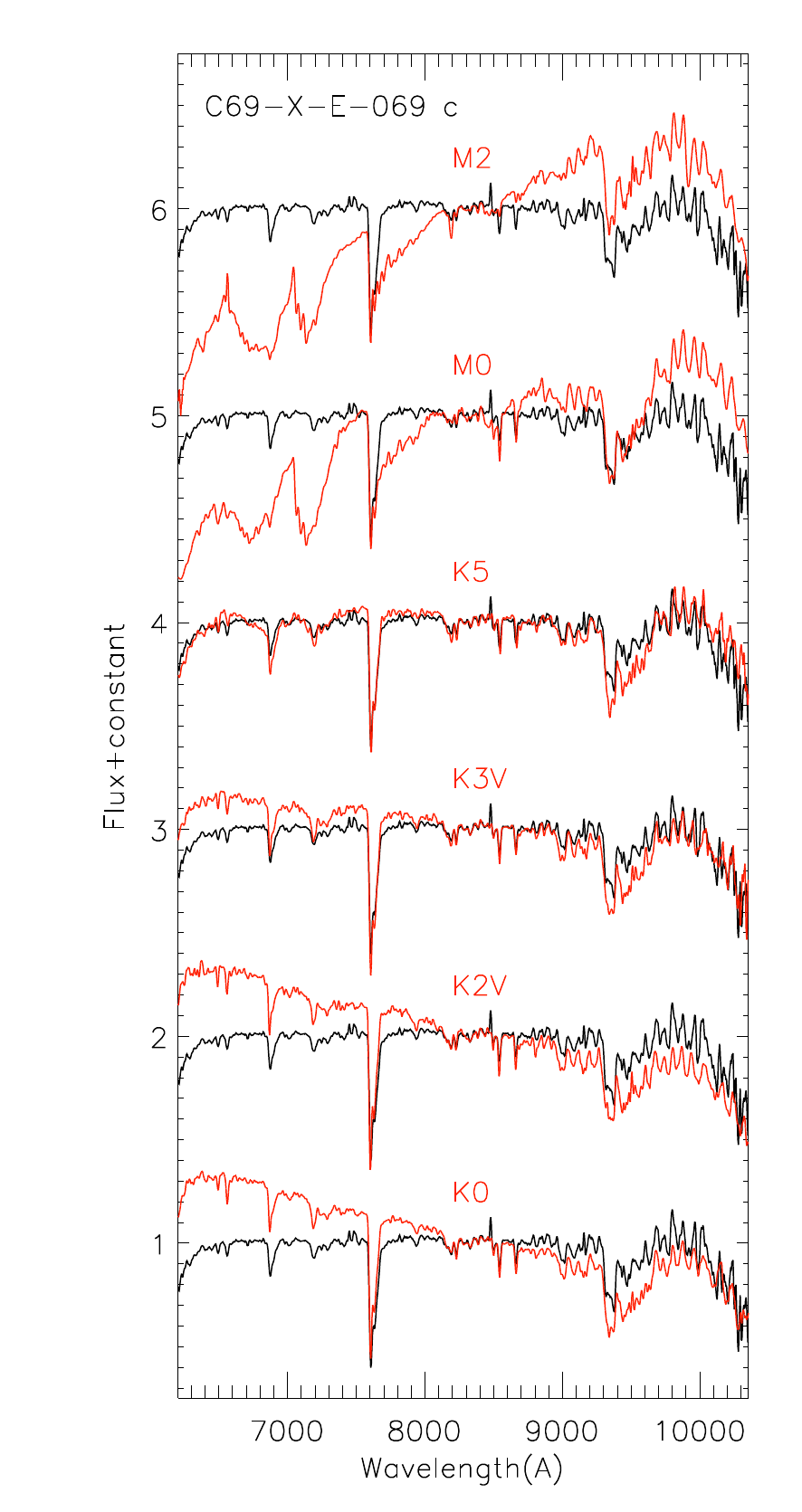}
\includegraphics[width=4.46cm, clip]{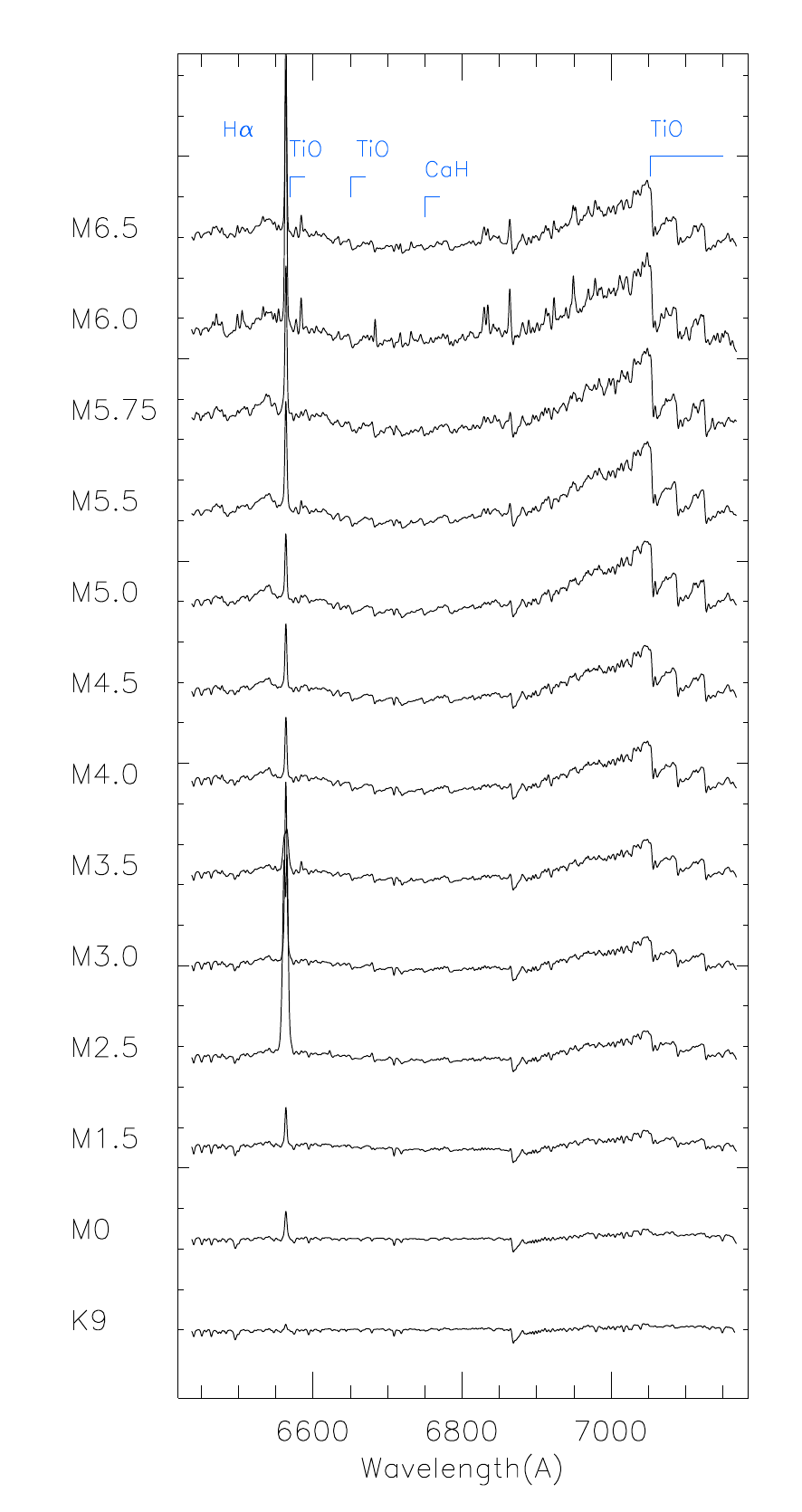}
\caption{{\bf Left:} Spectral type determination of one XMM candidate member by comparison to templates obtained with the same instrumental set-up (CAFOS). The ``science spectra'' are displayed in black and the templates (labelled according to the spectral type) in red. {\bf Right:} Spectral sequence derived for candidate members of Collinder 69 observed with FLAMES. Note the intense H$\alpha$ emission and strengthening of the typical TiO and CaH bands with the spectral type.}
\label{spt_derive}
\end{figure}

For the warmer part of the sample (a fraction of the XMM candidates), we compared our optical (low resolution, CAFOS) spectra with templates (obtained with the same configuration during the same campaigns) from Taurus members and field dwarfs within a spectral type range from G0 to M0 (the comparison spectra were reddened to the previously mentioned average low extinction, A$_{\rm v}\sim$ estimated by \citealt{Duerr82} for Collinder 69). We normalized both, the science spectra and the templates, at the same wavelength and by a simple $\chi^2$ minimization decided which template reproduced the science data best. We did not use specific temperature sensitive lines since our resolution did not allow us to fit anything else than the continuum shape. As an example, in Fig.~\ref{spt_derive} we show an example of this comparison for a K candidate member.

On the other hand, the majority of our candidates belongs to the colder sample, in principle with T$_{\rm eff}$ consistent with M spectral types. M-dwarfs are characterized spectroscopically by the presence of molecular bands of titanium and vanadium oxide (TiO, VO). These bands reach their maximum strength around M7, and then become weaker for cooler temperatures because of Ti and V condensation into dust grains. 
Regarding atomic lines, the Ca II triplet is much weaker than in M supergiant spectra, but on the other hand, the Na I and K I doublets are stronger, since they are gravity dependent. 

Several spectral ratios or indices (quantifying different band strengths) have been proposed in the literature to derive spectral types for these late-type stars. Some are based on the relative depths of the mentioned molecular bands
 (see for example \citealt{Reid95,Cruz02} and references therein); others are based on measuring the slope of the pseudocontinuum 
 (for example PC3 and PC6 from \citealt{Martin96} and \citealt{Martin99}).

We have used different combinations of these indices (depending on the resolution and wavelength coverage of the spectra) to classify our M-like candidates. In Fig.~\ref{spt_derive} we show an example of a spectral sequence obtained applying these indices to 13 candidates observed with FLAMES. This combination of indices should provide us with a spectral classification with $\sim$0.5 subclass accuracy.  

\begin{figure}[htbp]
\begin{center}
\includegraphics[width=4.4cm]{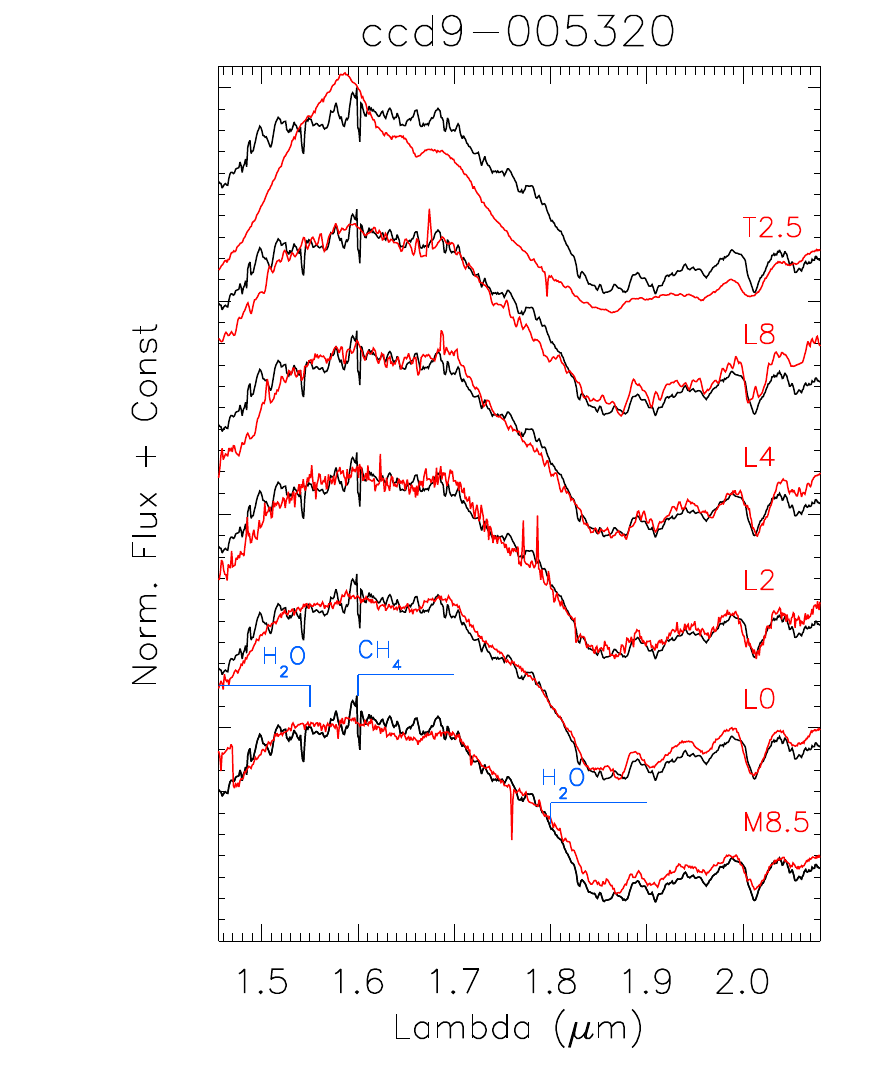}
\includegraphics[width=4.4cm]{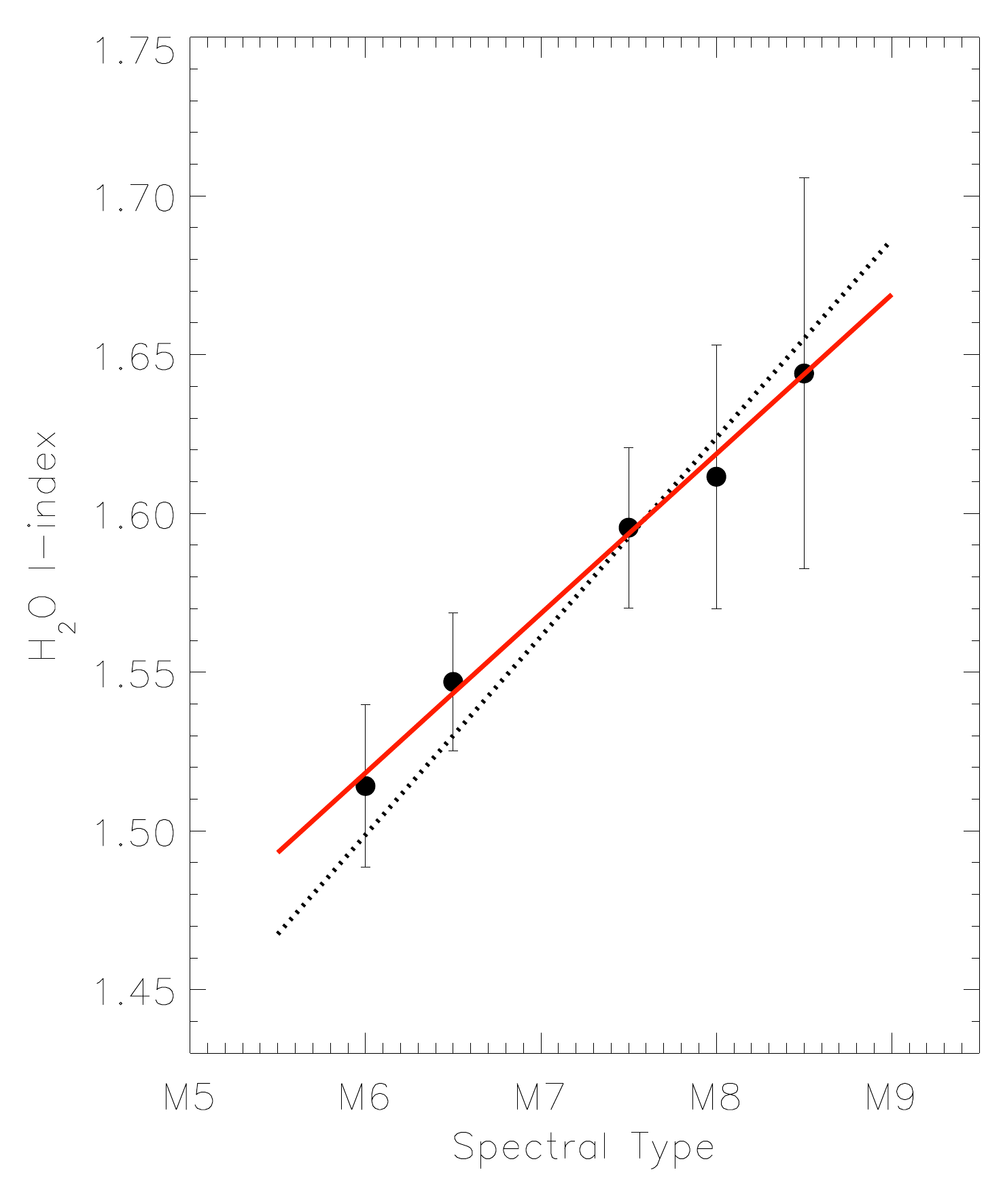}
\caption{{\bf Left:} Infra-red spectral type determination for a candidate member observed with Subaru. The templates used for comparison (red spectra with corresponding spectral type label) were also obtained with Subaru and correspond to known L and T dwarfs from \citealt{Geballe02}. A visual comparison with the spectra obtained by \cite{Lodieu08} of Upper Sco members also suggests L spectral type as a better match specially based on the water bands and the ``peaky" structure $\sim$1.7$\mu$m. {\bf Right:} Our own re-calibration of the  reddening independent $I_{\rm H_{2}O}$ index from \citet{Comeron00}. The dotted black line represents the linear relation found by \citet{Comeron00}, while the red line is the best linear fit to our data and is the relation that we have used to estimate spectral types of our M candidates.}
\label{fig:TspecFeat}
\end{center}
\end{figure}

{\bf Near-infrared spectra:} 
\label{IR_SpT}
Because of their nature, the reddest--coldest candidates of our sample (late M, L and even T spectral types according to the SED fitted temperatures), can be better studied in the infrared than in the optical. As in the optical case, we performed the spectral classification either through spectral indices (for the two M candidates, see right panel of Fig.~\ref{fig:TspecFeat}) or by comparison with templates (for the L and T candidates, see left panel of Fig.~\ref{fig:TspecFeat}).

The near-infrared spectra of M dwarfs are dominated by deep broad absorption bands of H$_2$O (particularly at 1.4 and 1.85 $\mu$m). The fact that water vapour in the Earth's atmosphere also contributes with a substantial absorption at these wavelengths hampers the analysis of these bands to some extent. However, the higher temperatures in the stellar atmospheres mean that the associated steam bands are broader than the terrestrial absorption, so the wings are accessible for measurement and analysis. 

Apart from these water bands, other molecular bands such as the CO (at 2.29 $\mu$m), FeH (at 0.99 $\mu$m) and VO (at 1.2 $\mu$m) also scale with temperature \citep{Jones94}. Nevertheless, these changes are not as dramatic as the ones observed on the water bands and, therefore, we will focus on the former to derive the spectral types.
  
\citet{Comeron00} defined a reddening independent index, $I_{\rm H_{2}O}$, to measure the depth of the wings of the water band of late M dwarfs centered near 1.9 $\mu$m. \citet{Gomez02} tested this method by comparing the estimated spectral types with the $Q$ index defined by \citet{Wilking99} and obtained excellent agreement (see Fig.8 in their paper). This $Q$ index is reddening independent too and was defined in order to characterize the strength of the 1.7-2.1 $\mu$m and $\ge$ 2.4 $\mu$m water absorption bands. 

Because of the wavelength coverage of our observations, we only used the $I_{\rm H_{2}O}$ index to estimate infrared spectral types of the candidates: In the right panel of Fig.~\ref{fig:TspecFeat} (dotted line) we show the linear relation proposed by \citet{Comeron00} between this index and the spectral type (obtained for confirmed members of Chamaeleon I, $\sim$1 Myr old). We have re-calibrated this relation using a sample of well-known field M-dwarfs (filled black circles with error bars illustrating the dispersion in the measurements made on $\sim$3 objects per spectral type) obtaining a very similar slope (red line). With this comparison, we see how, in the M5.5--M9 spectral range, the $I_{\rm H_{2}O}$ index is not sensitive to the known age dependency on the water bands. Keeping this caveat of the age uncertainty in mind, this relationship allows us to estimate spectral types within 1.5 sub-type. For the two objects where we used this method we also had optical spectra. In both cases, the infrared spectral type is colder than the optical one but within the error bars. This might be indeed a result of the age dependency of the $I_{\rm H_{2}O}$ index. Finally both sources were classified as diskless based on their IRAC photometry, so the possibility of some excesses caused by the disk affecting our classification can be ruled out.

Both optical and near-infrared spectral types derived by us are listed in Table~\ref{tab:paramTOTAL}.

\subsection{SED fitting}
\label{sec:SED}

As can be inferred from Section~\ref{sec:phot}, the photometric surveys provided us with a very large but inhomogeneous dataset. To analyze these data in a homogeneous and automatic manner we used the tool VOSA \citep{Bayo08}. 

In short, the user provides VOSA with a table with photometry and the tool performs the following tasks in an almost automatic way (very little interaction is needed): it enlarges the wavelength coverage by looking for counterparts in different (VO-compliant) catalogues, it builds the SEDs for all the sources, it performs the SED fit using models \citep{Hauschildt99, Allard01, Chabrier00, Castelli97} of stellar and substellar photospheres (effective temperature, surface gravity and metallicity as free parameters), it uses those models to obtain a multi-color bolometric correction and, finally, it interpolates among isochrones and evolutionary tracks \citep{Baraffe98,Baraffe02,Chabrier00,Baraffe03} to derive ages and masses.

In addition to the photometric data, the user has to provide VOSA with the estimated distance to each source and the extinction affecting the line of sight. In the case of Collinder 69, this does not represent any extra complication, since members should be located approximately at the distance derived for the cluster (400 pc, \citealt{Murdin77}) and  we benefit from the low extinction (A$_{\rm v}$$\sim$0.36 mag, \citealt{Duerr82}) affecting this region.

The parameters derived in this fashion are listed in Table~\ref{tab:paramTOTAL}. The relation between this VO methodology and other classical ones can be found in \citet{Bayo08}. In particular, some of the possible caveats to take into account in the determination of T$_{\rm eff}$, L$_{\rm bol}$ and Mass via SED fit are:

- Blue excess in the SED due to accretion. This aspect is discussed in Paper II: there we study in detail the veiling in the objects with the strongest H$\alpha$ emission and demonstrate that for these cool objects veiling is hardly noticeable and therefore the determination of T$_{\rm eff}$, L$_{\rm bol}$ and Mass via SED fit is not sensitive to it.

\begin{figure}
\includegraphics[width=9cm, clip]{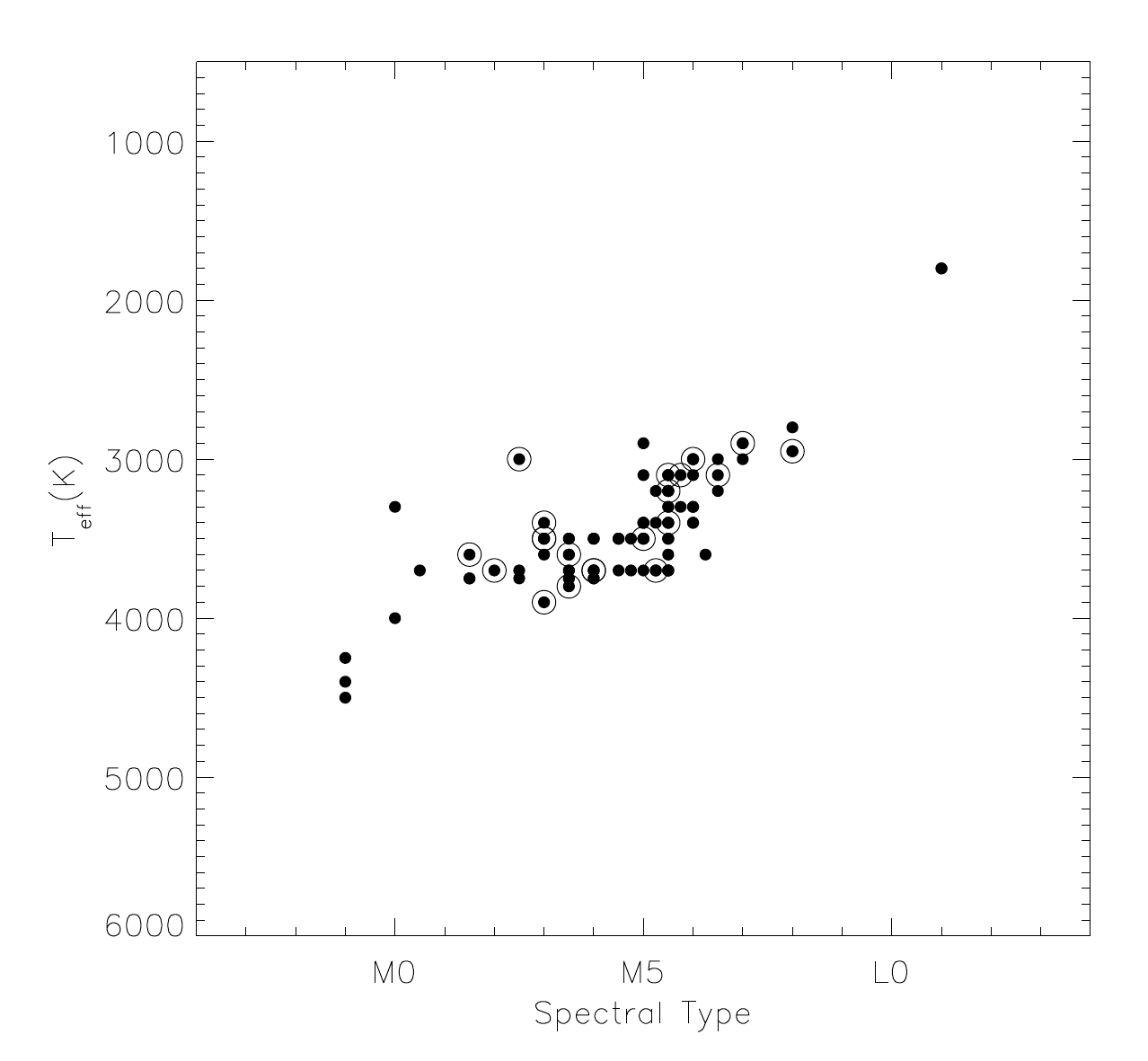}
\caption{Comparison between the spectral types determine from spectra and the effective temperature estimated via SED fit. We only display the spectroscopically confirmed members here, highlighting the sources that harbour disks with large circumferences. It is clear that no systematic differences in T$_{\rm eff}$ estimation can be attributed to the presence of disks.}
\label{teff_spt}
\end{figure}

- Red excess in the SED due to the presence of a circumstellar disk. The impact of circumstellar disks on our SED fits is discussed in \cite{Bayo08}, where we showed that only edge-on disks above a certain mass (below which, the disk is not massive enough to produce an effect in the photospheric part of the SED) introduce a bias in our parameter estimations.
When this is the case, our methodology will underestimate both the effective temperature and the bolometric luminosity of the central object. 
From the SED shape of the confirmed members (no infrared peak higher than the photospheric one) we can infer that we do not have edge-on disks in our sample. In any case, to illustrate this point more clearly, in Fig.~\ref{teff_spt} we plot the estimated spectral type vs effective temperature. While the latter could in principle be affected of the presence of the disk, the former should not since we are using blue features of the optical spectra to estimate the spectral type. In the figure we have highlighted the objects that harbour disks by plotting large circles around them. We can see how the dispersion in effective temperature among the objects of the same spectral type does not correlate with the presence of disks.

- Gray excess in the SED due to multiplicity. To study the effect of possible unresolved multiple systems we have performed a simple exercise considering two cases: a similar mass ratio and an ``extreme" one. In both cases we have built simulated SEDs by adding the fluxes of two pairs of confirmed members and performed the fit in the composite SED. As expected, when the components have similar characteristics (in this particular case M4 spectral type, diskless sources); the derived effective temperature does not change, but VOSA over-estimates the mass and L$_{\rm bol}$ of the individual components (and of course, underestimates the age). On the other hand, for the most extreme case that we could build (class III sources but with 5000 and 2800 K effective temperatures), the resulting T$_{\rm eff}$ is very close to that of the hottest component (4700 K) and the L$_{\rm bol}$ larger than that estimated for the hottest component, but still almost within the error-bars. Thus, in this case again, the larger effect to keep in mind is that the age of  the system would be underestimated.
In any case, according to \cite{Sacco08} and \cite{Maxted08}, the, -short period-, binarity fraction in Collinder 69 is as low as $\sim$10\% (unlike the higher $\sim$30\% reported for field M dwarfs by \citealt{Reid97}); so, although for the individual targets we should keep these possible biases in mind; the general conclusions about the cluster as a whole should not be affected by binarity.

\subsection{Effective temperature scale.}
\label{sec:teffScale}

Several temperature scales for M dwarfs are already available in the literature \citep{Bessell98,Luhman99, Leggett00, Luhman03}. However, in many cases, the sample of objects for which this scale is derived lacks homogeneity and/or is based on small numbers statistics. As an example, the set of objects studied sometimes comprises sources with different ages or environments, and therefore the possible effects of those factors cannot be addressed.

\begin{figure}
\includegraphics[angle=0,width=9cm]{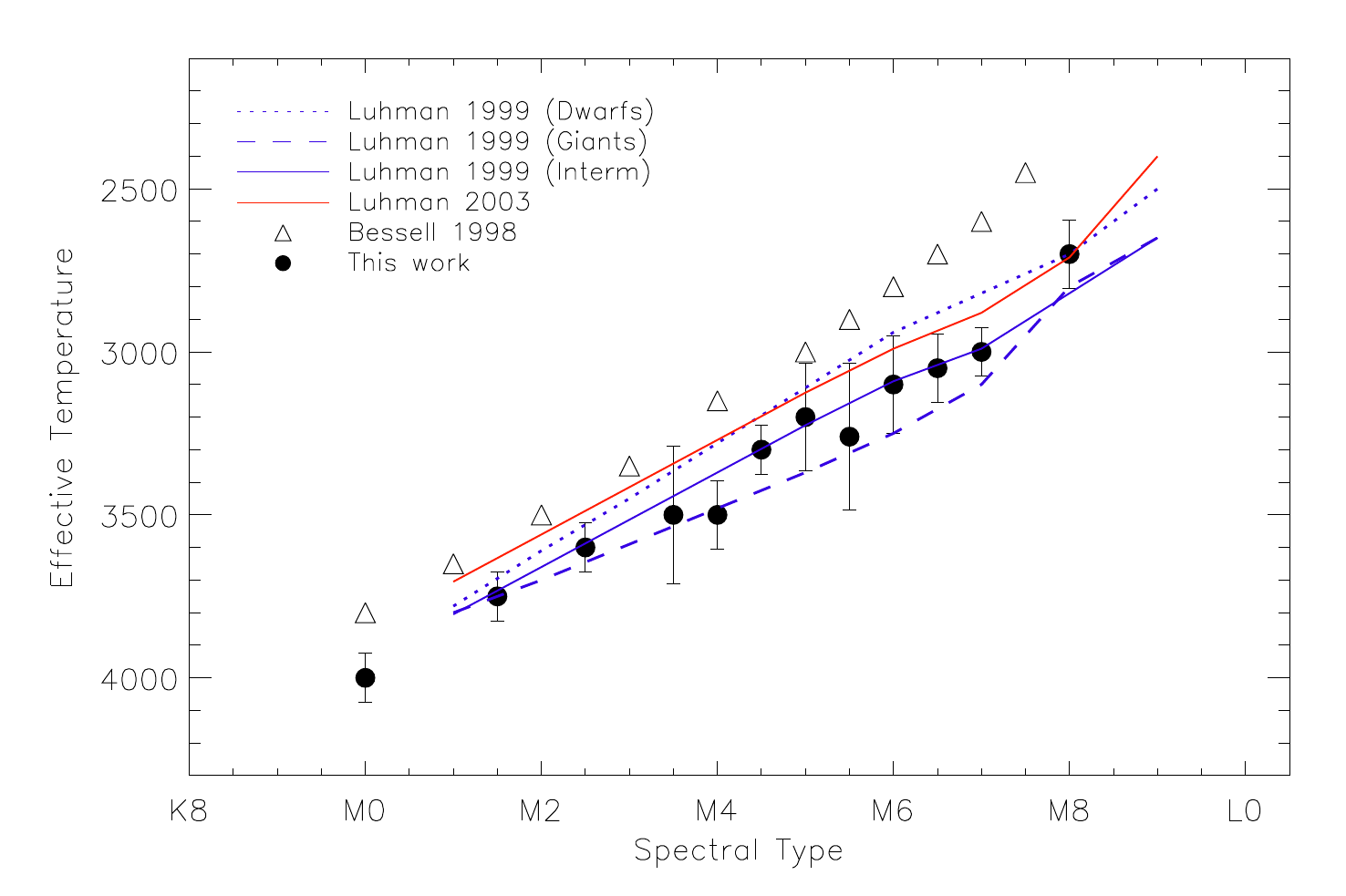}
\caption{Temperature scale derived for the confirmed very low mass M star and brown dwarf members of Collinder 69. With different symbols and line styles we also display scales previously reported in the literature.}\label{fig:teffScale}
\end{figure}

With this in mind, using the determinations of spectral types and effective temperatures from the previous two subsections, we have derived our own temperature scale. We are taking advantage of the homogeneity of our determination of spectral types and effective temperature and of the fact that we are studying objects with roughly the same age, that were born in the same environment (an environment with the practical advantage of low extinction affecting the observational data). To derive the temperature scale, we started by defining a ``clean" sample (see Table~\ref{tab:teffScale} and Fig.~\ref{fig:teffScale}). This ``clean" sample is composed of $\sim$30 sources fulfilling every one of the next criteria: membership confirmed through lithium absorption (see next section), SED with no infrared excess whatsoever (to avoid problems in T$_{\rm eff}$ determination implied by edge-on disks with VOSA, see Section~\ref{sec:SED} and \citealt{Bayo08}), optical spectral type derived between M0 and M8 (see Section~\ref{sec:SpT}), no signs of variability or binarity and estimated age according to the HR diagram between 1 and 10 Myr (see Fig.~\ref{fig:HR}). 

The temperature scale determined in this manner is shown in Table~\ref{tab:teffScale}. In Fig.~\ref{fig:teffScale} we compare our scale with others from the literature \citep{Luhman99, Bessell98,Luhman03}. We find a good agreement with the scale derived by \citet{Luhman99} for a sample of Taurus members (1--3 Myr, slightly younger than Collinder 69) but systematically higher temperatures than those proposed by \cite{Bessell98} and \cite{Luhman03}. With respect to the \cite{Bessell98} scale; what we see is probably just the effect of the difference in age from the two samples of objects for which the scales are derived. 

On the other hand, the scale presented by \cite{Luhman03} correspond to an update of the one derived in \cite{Luhman99}, but in this case for the young cluster IC348 ($\sim$2 Myr old). Therefore, the differences between the two scales in this case should not be related to an aging effect. In any case, these differences are not significant taking into account the dispersion in estimated T$_{\rm eff}$ that we see for an individual subclass (specially between M3.5 and M6). 

In this same line, we must note that the dispersion previously mentioned, seems to be of the same order as the differences found by \citet{Luhman99} between dwarfs and giants (while none of our confirmed members can have a low luminosity class according to our study). This dispersion seems to be higher within specific spectral types and might be revealing some physical properties of the atmospheres of objects within this range of temperatures. It could be related to dust settling, although the temperature range seems to be too high for this phenomenon to happen. It could be related to metallicity, but, to the best of our knowledge, there are no detailed metallicity studies in this cluster. In any case, our ``clean" sample is not well enough sampled to allow us to derive any conclusion.

\begin{table}
\caption[Temperature scale derived from our spectroscopic data of Collinder 69.]{Temperature scale derived from our spectroscopic data of Collinder 69.}
\centering
\label{tab:teffScale}
\begin{small}
\begin{tabular}{cc}
\hline\hline
Spectral Type & Effective Temperature\\
\hline
M0   &    4000 K \\
M1.5   &    3750 K \\
M2.5  &    3600 K \\
M3.5  &    3500 K \\
M4   &    3500 K \\
M4.5 &  3300 K\\
M5   &    3200 K \\
M5.5 & 3260 K\\
M6   &    3100 K \\
M6.5 & 3050 K\\
M7   &    3000 K \\
M8   &    2700 K \\
\hline\hline
\end{tabular}
\end{small}
\end{table}

\section{Confirming membership}
\label{sec:membership}

To confirm the membership of our candidates we have used different diagnostics (all of them based on alkali absorption lines) depending mainly on the resolution of the spectra:

\subsection{Lithium absorption at 6708 \AA:}
Above $\sim$0.065 M$_{\odot}$ (less massive objects cannot develop the necessary temperature in their cores), lithium acts as an age scale because the time it takes for the core to reach 3.0$\times$10$^6$ K is a sensitive function of mass \citep{Basri97}. Very low mass stars and brown dwarfs down to this mass limit are fully convective, and therefore, once the core temperature exceeds the necessary limit, the entire lithium content of the star should be exhausted rapidly and thus be reflected in an observable change in the photospheric lithium abundance.Theoretical models make specific predictions about the time evolution of this lithium depletion boundary. For example, \citet{Ventura98} predict that at ages 30, 70, and 140 Myr, the lithium depletion edge should occur at 0.17, 0.09, and 0.07 M$_{\odot}$, respectively. Other models by \citet{Chabrier97} and \citet{Burrows97} make similar predictions of the variation of this lithium depletion boundary with age. Indeed, \citet{Bildsten97} and others have argued that the age for an open cluster derived in this manner should be more accurate than those found by any other method \citep{Stauffer98, Stauffer99, Barrado99, Jeffries03,Barrado04,Jeffries05,Jeffries09}.

\begin{figure}[htbp]
\begin{center}
\includegraphics[width=9.cm]{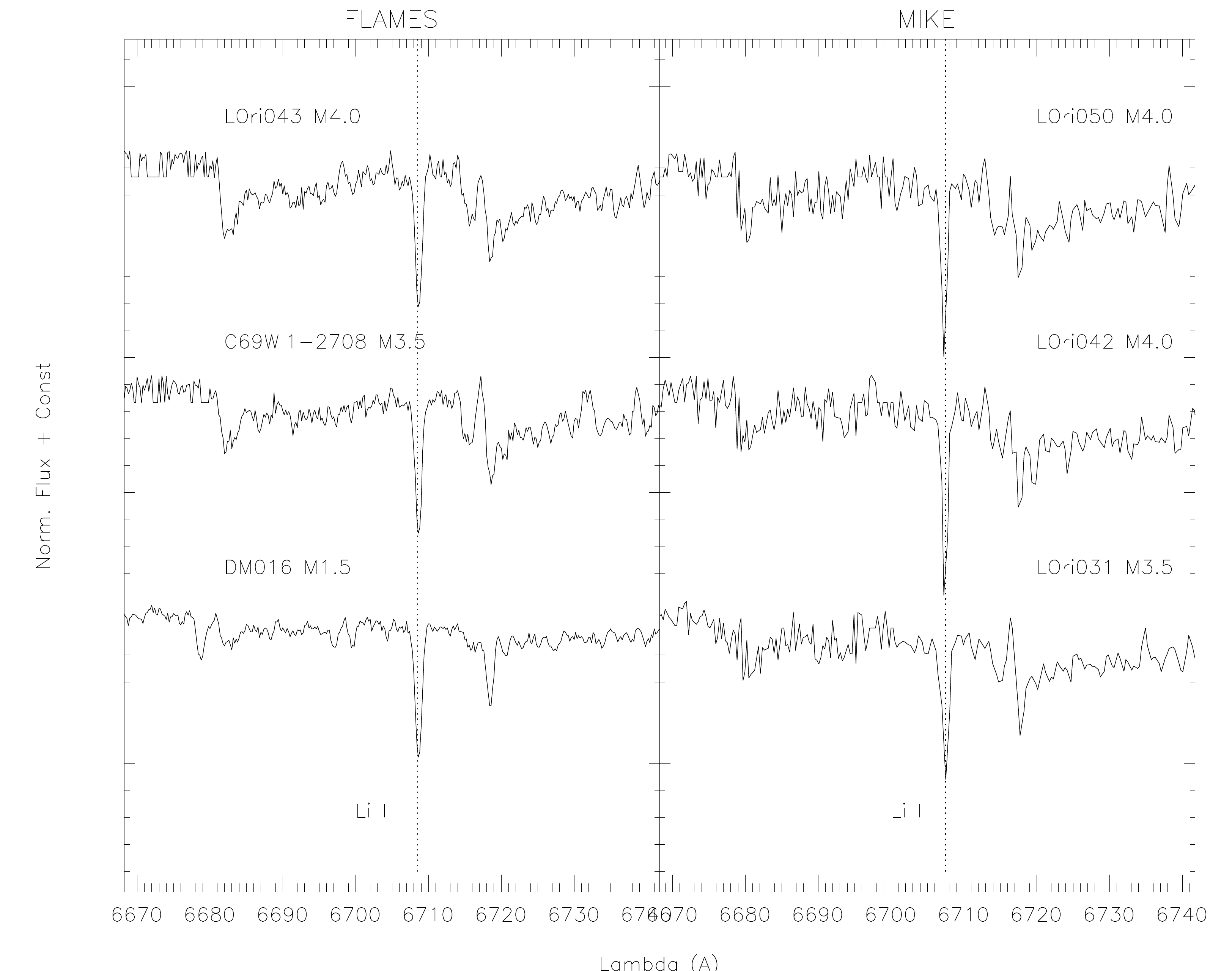}
\caption[Detail around Li I $\lambda$6708\AA.]{Detail around  Li I $\lambda$6708\AA~for eight confirmed members with different spectral types. The two panels correspond to two different campaigns: Magellan/MIKE echelle spectra (only the order corresponding to this line), and FLAMES multi-fiber spectra.}
\label{C69:LiI}
\end{center}
\end{figure}

Because of the intrinsic weakness of the lithium line we have only been able to use it as a youth indicator for the sources observed with a high signal-to-noise ratio (especially in the lower resolution campaigns) and down to a certain spectral resolution (see Appendix A). Whenever these two factors were favorable, the lithium presence (and absence) was considered the main component to assess membership in Table~\ref{tab:paramTOTAL}. As an example, Fig.~\ref{C69:LiI} shows two objects with clear Li I absorption in spectra taken with different instrumental set-ups. 

\begin{figure}[htbp]
\begin{center}
\includegraphics[width=9.4cm]{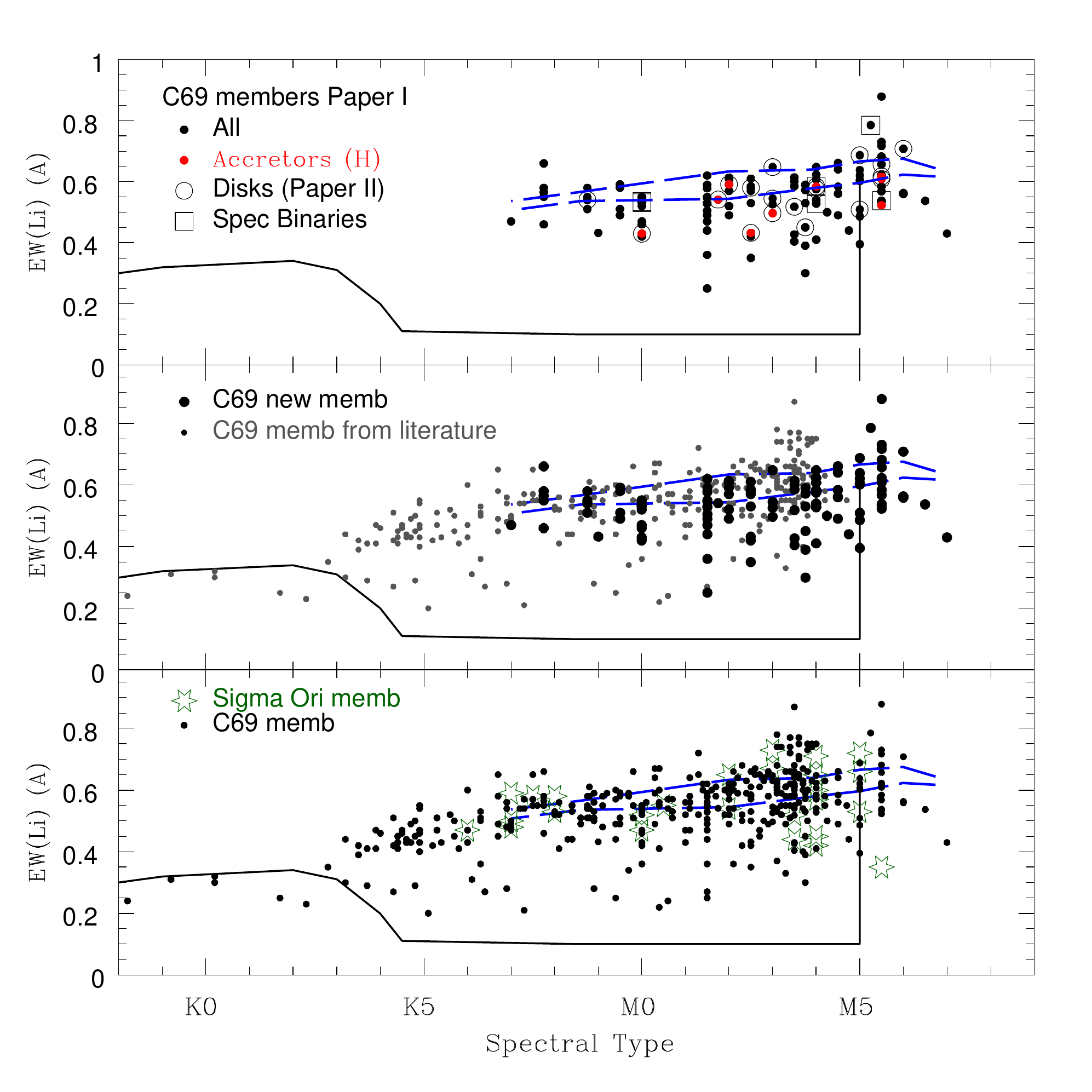}
\caption{Lithium equivalent width (EW, in~\AA) versus the spectral type for different sets of spectroscopically confirmed members of Collinder 69. {\bf Top:} New members reported in this work where accretors (based on the H$\alpha$ emission, see Paper II for details) are highlighted in red, and Class II sources (based on IRAC data, see \citealt{Barrado07} and Paper II) and spectroscopic binaries (from \citealt{Sacco08} and \citealt{Maxted08}) are surrounded by open circles and squares, respectively. {\bf Middle:} New members showed as filled black circles, while members from the literature \citep{DM99, DM01, Barrado04, Barrado07} are shown as smaller grey circles. {\bf Bottom:} Comparison of Collinder 69 members (joining the samples from the middle panel), displayed as filled black dots, with Sigma Orionis low mass stars from \citet{Zapatero02}, represented with star symbols. Regarding all the panels: The solid line traces the upper envelope of the values measured in older clusters such as IC2391, IC2602, the Pleiades and M35;  and the short-dashed blue lines corresponds to the cosmic abundances -- A(Li) = 3.1 --  from gravities of log g = 4.5 and 4.0, respectively (curves of growth from \citealt{Zapatero02}).}
\label{C69:LivsSpT}
\end{center}
\end{figure}

To illustrate the presence of Li I in absorption as a youth indicator, Fig.~\ref{C69:LivsSpT} shows the measured lithium equivalent width versus the spectral type of the (newly confirmed) members to Collinder 69. In the middle panel, we have included the data from \citet{DM99, DM01, Sacco08} along with the measurements from this work. Whenever more than one measurement of the Li equivalent width was available we display the average value. For comparison, in the bottom panel of the same figure, we have included data corresponding to a similar age cluster, $\sigma$ Orionis, as well as (in all the panels) an upper envelope of the values measured in older clusters (see caption of the figure). We have also plotted the theoretical equivalent widths from \citet{Zapatero02} considering $\log g = 4.0$ and $\log g = 4.5$ and initial cosmic abundance (A(Li) = 3.1). 

The scatter of the Li I equivalent widths is considerable for all spectral types included in the figure, getting even larger in the M domain. This is a well-known trend for which there is not a clear explanation yet (see \citealt{Barrado01}, \citealt{Zapatero02}, and more recently \citealt{daSilva09} and references therein).
The dispersion could be ascribed to a variability in the Li I line as a consequence of stellar activity, different mixing processes, presence or absence of circumstellar disks, binarity, or different rotation rates from star to star. Recently, \cite{Baraffe10} tried to explain this dispersion in terms of the early accretion history that could lead to a dispersion in the Li abundances of young low mass stars and brown dwarfs. In the top panel of Fig.~\ref{C69:LivsSpT}, we have plotted red dots on top of the sources in our sample classified as accretors (based on the H$\alpha$ equivalent width, see Bayo et al 2011b, hereafter Paper II, for details), large open circles around those showing infrared excess and large open squares for those sources classified as binaries by either \cite{Sacco08} or \cite{Maxted08}. It turns out to be quite clear that none of these special sets of objects show distinct positions in the diagram. Therefore the cause for the scatter observed in these measurements remains unknown.

Based on this scatter, \cite{daSilva09} warned about the possible problems of using Li to date individual stars; we have tried to quantify whether this can be a problem in our case, and we have found that in all the cases (16 sources) where membership was confirmed by \cite{DM01} or this work based on Li and where \citet{Maxted08} had studied the radial velocity; membership was confirmed by them too. On the other hand, we do not have means to measure the opposite effect, since for the only two objects that we have discarded as members based merely on lithium (LOri011 and LOri012), there is no radial velocity measurement to compare with. In any case, for very young associations like this one, and for the estimated spectral types of these two objects (K8 and K9), even with scatter, lithium should be detectable (see Fig. 2 of \citealt{daSilva09} for more details).

\begin{figure}[htbp]
\begin{center}
\includegraphics[width=8.5cm]{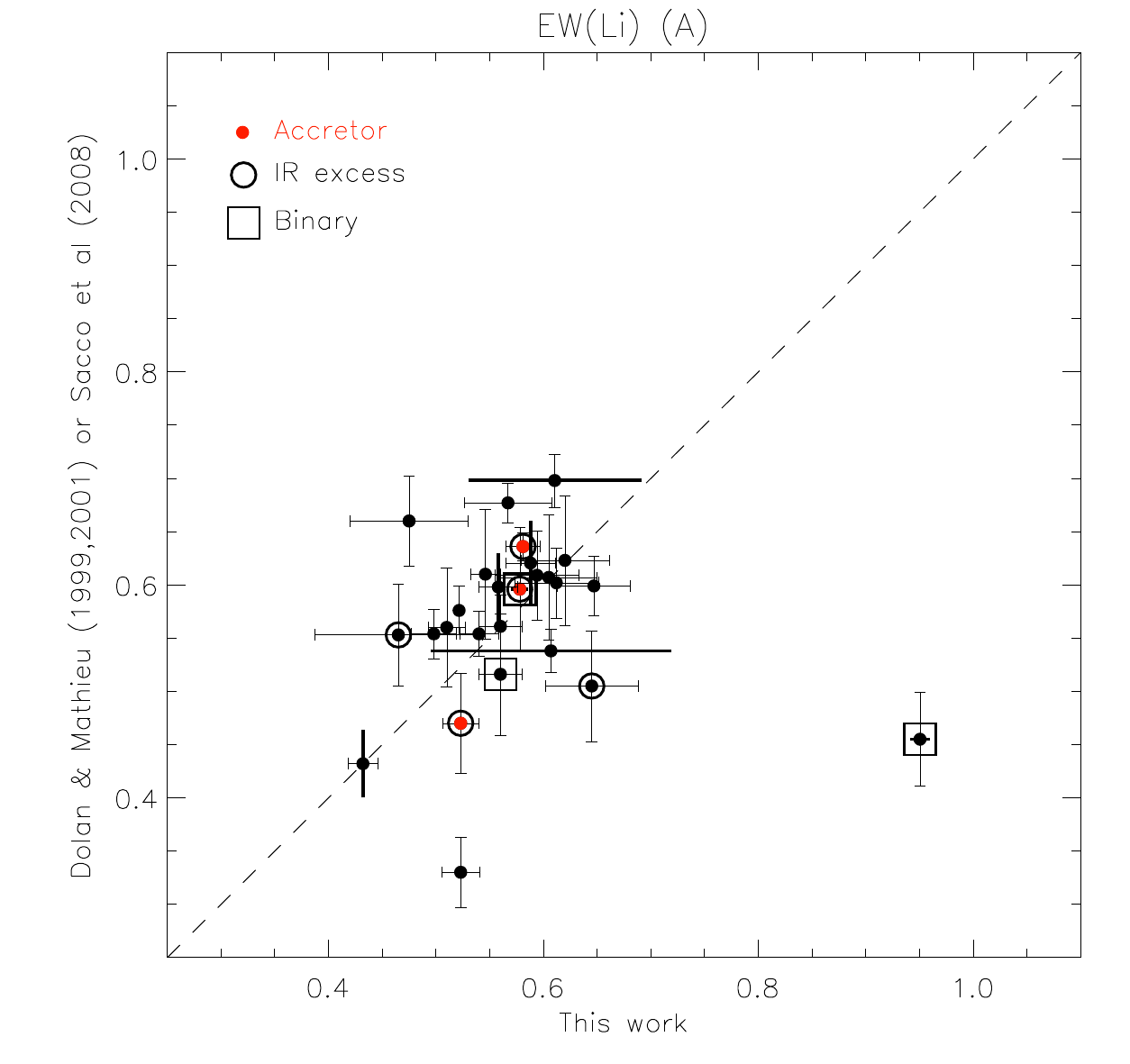}
\caption{Lithium equivalent width listed in \cite{Sacco08} or \cite{DM99,DM01} versus the values measured in this work, in~\AA. We highlight especial sources in red (accretors), or surrounding them with a circumference (objects showing infrared excess) or a square (binary systems according to \citealt{Sacco08, Maxted08}). Vertical or horizontal solid lines represent the range in variation when more than one measurement existent for one of the axes.
}
\label{C69:Li_var_fig}
\end{center}
\end{figure}

In addition to the dispersion in the lithium EWs among sources with different temperatures, for some objects we have observed some variability in this line when comparing our data with the values measured by \cite{Sacco08} or \cite{DM99, DM01} (see also Sec 5 on Paper II). We display our measurements against theirs in Fig.~\ref{C69:Li_var_fig}. It looks like most of our values are in good agreement (within the error-bars) with those of the literature. However, for the sources listed in Table~\ref{tab:Li_var}, the differences cannot be explained in terms of uncertainty of the measurements. Most of these objects show features typical of active stars (such as X-ray emission or H$\alpha$ variability). \cite{Neuhaeuser98} monitored the young star Par 1724 finding a variable lithium equivalent width consistent with rotational modulation. In any case we only have two measurements per source, and the sample is too small to reach any conclusion.

\begin{table*}[]
\begin{center}
\caption{Properties of the sources showing lithium EW variability (\citealt{Sacco08,DM99,DM01} values compared to the values derived in this work, third column). Note that there is no bias concerning the resolution of the spectra (as was also proved with the exercise of Appendix A). The instrument code is the same as in Table~\ref{tab:paramTOTAL}.}
\label{tab:Li_var}

\begin{tabular}{lllllrlll}
\hline\hline
Object & SpT & INS & Binarity$^1$ & IR Class & T$_{\rm eff}$$^2$ & vsini$^3$ & H$\alpha$ Vble$^4$ & X-ray$^5$ \\
\hline
LOri013 & M3.5 & (8) & & III & 3750 &-- &-- & Y\\
LOri045 & M3 & (8) (6)& &III&3500&$<$17&Y&N\\
LOri057 & M5.5 & (3) &  & III & 3700 & $<$17.0 & N & N \\
LOri063 & M4.0 & (3) &  & II & 3700 & $<$18.2 & Y & N\\
LOri068 & M4.5  M5.0 & (1)  (3) &  & III & 3700.0 & $<$17.0 & Y & N\\
LOri075 & M5.0  M5.5 & (2)  (8)  (3) & SB1 (M08) & III & 3400 & $61.3^{+11.5}_{-4.9}$ & Y & N\\
LOri088 & M5.5 & (8) &  & III & 3200 & $<$17.0 & Y & N\\
LOri094 & M5.5 & (8)  (1) &  & III & 3200.0 & $54.8^{+5.5}_{-8.2}$ & Y & N\\
LOri106 & M5.5 & (1) &  & II & 3200 & $<$17.0 & Y & N\\
\hline\hline
\end{tabular}
\end{center}

\begin{footnotesize}
$^{1}$ According to \cite{Sacco08} (S08) or \cite{Maxted08} (M08).\\
$^{2}$ Derived with VOSA.\\
$^3$ From \cite{Sacco08}.\\
$^4$ Comparing the measurements of this work (see Paper II) and/or with those of \cite{Sacco08}.\\
$^5$ Sources detected in our {\it XMM-Newton} survey (see Section 2.1).\\
\end{footnotesize}
\end{table*}

\subsection{Sodium and potassium absorptions:}

Regarding the lower resolution sample of sources, other alkali lines (more prominent than Li I) are known to be gravity sensitive in M-type stars. This is the case for some sodium and potassium doublets: K I at 7665 \& 7700 \AA~ and Na I at 8184 \& 8195.5  in the optical and K I at 1.169 \& 1.177 $\mu$m and 1.244 \& 1.253 $\mu$m in the near-infrared (see \citealt{Martin96, Schiavon97}). The surface gravity of members of Collinder 69 in this range of temperatures is expected to be $\log(g) = 3 - 4$ \citep{Baraffe98}, whereas a typical M-type giant will have $\log(g) \sim 2$ and an older main sequence dwarf will have $\log(g) \sim 5$. Thus, the equivalent width of these doublets can be used to identify (and discard as members) background giants and foreground dwarfs from our sample. 

To illustrate these differences, in Fig.~\ref{fig:NaIno-memb}, we show a comparison of the Na I doublet of candidate sources with field dwarfs of the same spectral types.

Due to the inhomogeneous nature our data, we had to chose the most suitable lines to be compared with templates or models depending on the resolution and the wavelength coverage of each campaign. 
As an example, for the medium resolution LRIS sample, we have only been able to study the bluest component of the optical K I doublet (the wavelength coverage of the instrumental set-up did not reach the other component).
We must note anyway that the EW of this component measured on the templates has a very strong dependence on the luminosity class, and therefore we think that our estimated values listed in Table~\ref{tab:paramTOTAL} are reliable. 

\begin{figure*}[ht]
\begin{center}
\includegraphics[width=15.cm]{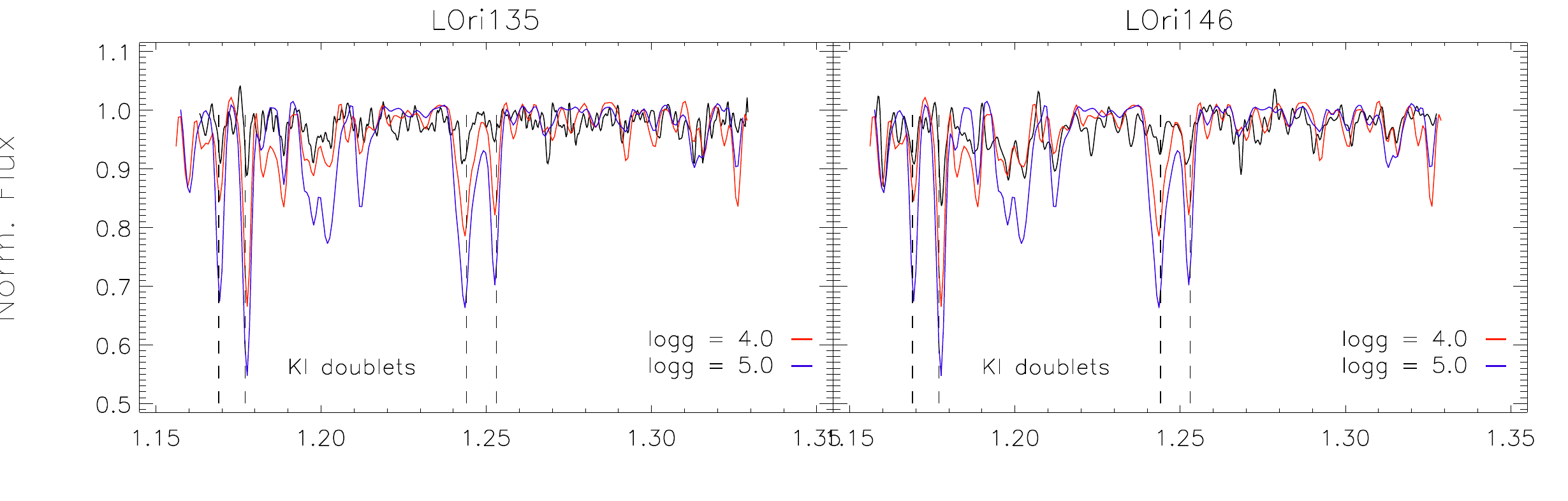}
\caption{Example of the surface gravity estimation for two NIRSPEC spectra by comparison with theoretical models.}
\label{fig:logg_determ}
\end{center}
\end{figure*}

As mentioned before, we have also used both, templates and models \citep{Allard03}, to assess membership. In Fig~\ref{fig:logg_determ} we show the cases of LOri135 and LOri146: with estimated temperatures of 3000 and 2800 K respectively, we have normalized both: the science spectra, and models for those temperatures and different values of $\log(g)$. Even though the signal to noise ratio was not too high, it is very clear that these sources must have a value of $\log(g)$ lower than 4.0 (and higher than giant-like values, where these doublets are barely detectable).

\begin{figure}[htbp]
\begin{center}
\includegraphics[width=8.7cm]{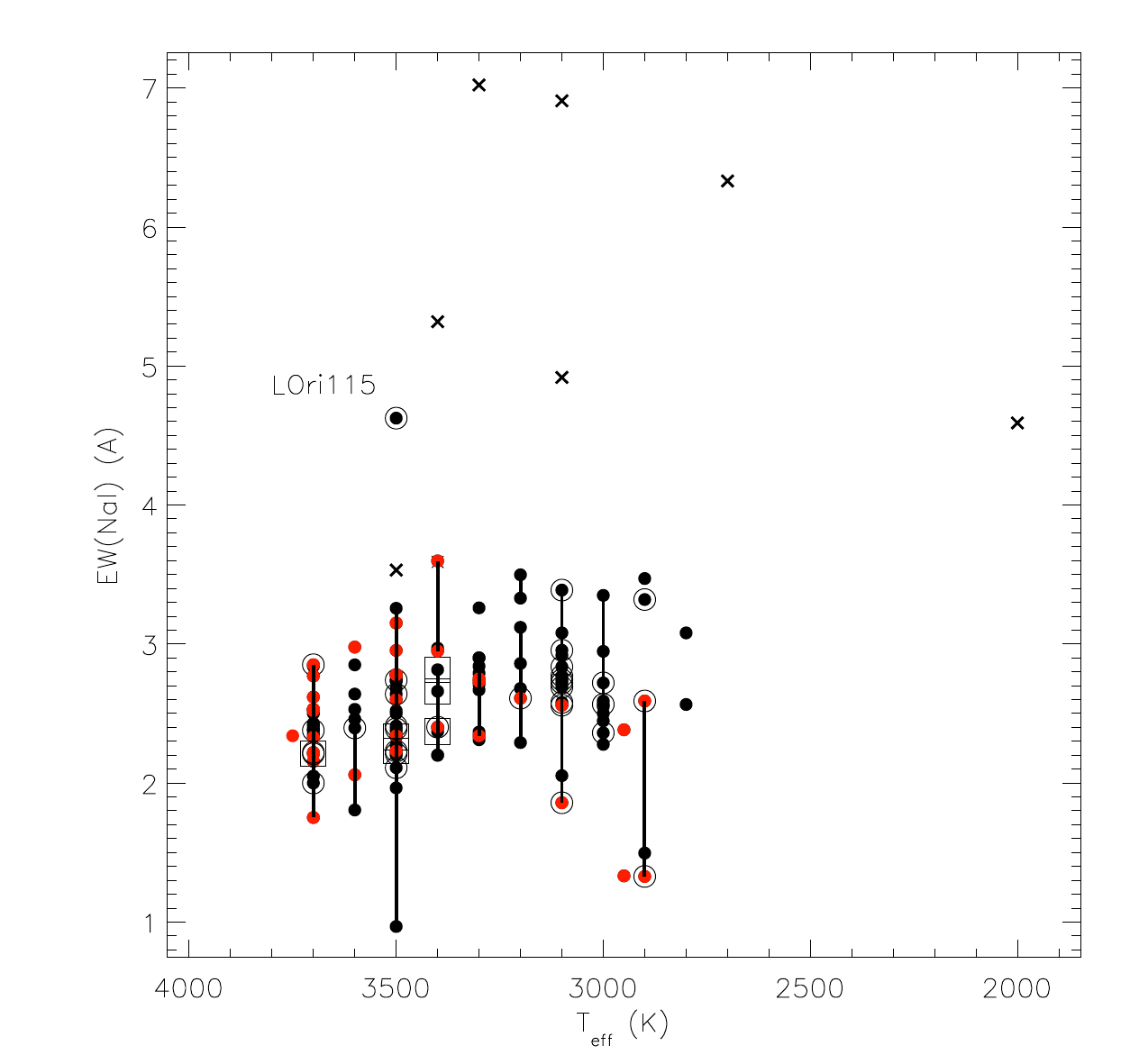}
\caption{T$_{\rm eff}$ vs EW(NaI) for members and non-members of Collinder 69. Rejected candidates are displayed as crosses, confirmed members as solid dots, where: red ones highlight sources classified as active accretors and/or detected on our X-rays survey. Large open circles and squares surround sources showing infrared excess or classified as binaries in either \cite{Sacco08} or \cite{Maxted08} . For those sources showing large variability in the measured Na I equivalent widths (measurements from \citealt{Maxted08} and this work) we have included vertical bars connecting those measurements. Note the unusual position of LOri115 (discussed in Section~\ref{PS:memb})}
\label{C69:Teff_NaI}
\end{center}
\end{figure}

In Fig.~\ref{C69:Teff_NaI} we plot the effective temperature (in the range where the method is applicable, see Sec.~\ref{PS:memb}) of members and rejected candidates to Collinder 69 against their measured Na I EW (the $\lambda$8200 doublet). We have also included the measurements from \cite{Maxted08} and highlighted in red those sources showing X-ray emission and/or sources that are classified as active accretors in Paper II. LOri115, a ``peculiar" source (see further on, in the text), is marked as well. As we did with the Li I measurements, we have compared the values measured by us with those from \cite{Maxted08} for the sources in common. We find that $\sim$35\% of the common sources show Na I variability (differences not compatible within the error bars for both measurements). The majority of these sources are either Class II sources or X-ray emitters and therefore this variability must be mainly related either to activity or accretion processes. 

This diagram may suggest also that in the substellar domain (the boundary is located at T$_{\rm eff}$$\sim$3150 K for 5 Myr according to the SIESS + COND isochrone) active accretors show lower Na I EWs than non-accreting brown dwarfs. A first explanation for this trend would be that our measurements of Na I EWs are affected by veiling for the accreting brown dwarfs; this is not the case, since, as it is described in Paper II, even for the sources with the most intense H$\alpha$ emission, the veiling is minimal for wavelengths redder than 7400 \AA. In any case, the sample is too small and the trend too shallow to allow us to perform any further analysis. Finally, the position of LOri115 in the previously mentioned diagram seems suspicious (the Class II very low mass star clearly out of the general trend of the confirmed members), but its membership is discussed in the following section.

\begin{figure}[htbp]
\begin{center}
\includegraphics[width=9.cm]{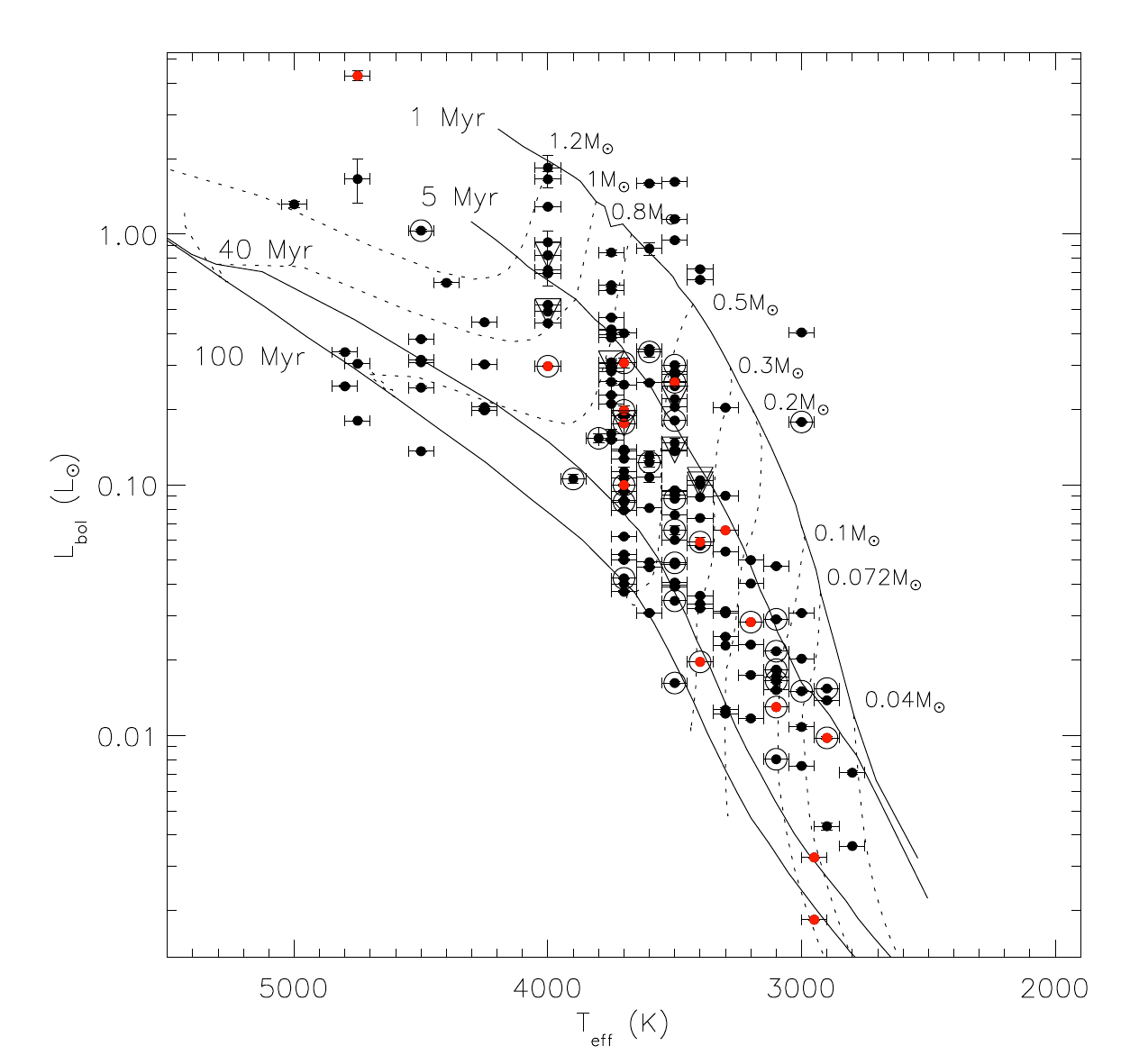}
\caption{HR diagram for the $\sim$200 spectroscopically confirmed members by \cite{DM99,DM01,Maxted08,Sacco08} or this work. We have over-plotted several isochrones and evolutionary tracks from \cite{Baraffe98}. Symbols as in Fig.~\ref{C69:Teff_NaI}}
\label{fig:HR}
\end{center}
\end{figure}

\subsection{Peculiar Sources:}
\label{PS:memb}

We distinguish in this section among four different groups: sources for which we have information about the Li absorption; objects for which we have information about other alkali lines (and/or emission lines and infrared excess); targets for which, due to the lack of high resolution spectra and its ``hot" nature, we have had to use different criteria than alkali lines to study membership; and, finally, the coldest sample of sources only observed with SUBARU:

{\bf 1) Sources with Li information:}

{\bf LOri044, LOri046, LOri049, LOri052:} All these sour\-ces (except LOri052) have been classified as non-members based on the absence or very marginal detection of Li I. In the cases of LOri044, LOri046, LOri049, LOri052, either \citet{Maxted08} or \citet{Sacco08} also classify them as non-members based on their radial velocities. In our FLAMES spectra of LOri052, a marginal detection of Li I in absorption with EW of $\sim$0.28 \AA~ was measured; this value is too low compared to other members with similar spectral type (even taking into account the general spread already commented on); therefore, together with the radial velocity determination of \citet{Maxted08} and our alkali lines study, we classify this source as probable non-member. 

{\bf LOri007:} No significant H$\alpha$ emission was measured on the spectra of this source (observed with TWIN and with FLAMES), and although the equivalent width values of the Na I doublet is consistent with membership (see Table~\ref{tab:paramTOTAL}), no Li I in absorption has been detected (see Table~\ref{tab:paramTOTAL}). Therefore we provide a classification of non-member.

{\bf DM065, DM070, DM061, LOri001, DM062, DM016, LOri026 (DM012), LOri038 (DM002):} with spectral types from K9 to M3: confirmed by \cite{DM99} via Li absorption. Besides, their positions in the HR diagram are compatible with membership and most of them are X-ray emitters.

{\bf LOri115}: Shows a deeper Na I absorption (LRIS spectrum) than expected for its spectral type (M5) but a Li EW (FLAMES spectrum) perfectly compatible with membership. Besides, the SED of LOri115, exhibits a clear infrared excess at IRAC wavelength suggesting that the source is surrounded by an optically thick disk. It also shows quite strong and variable H$\alpha$ emission (EW of -12.5 and -9 in our FLAMES and LRIS spectra), very close to the limit where pure activity cannot explain this emission. All this clear signs of youth allow us to classify it as member and propose as possible explanation for the deep Na I absorption some phenomenon related with activity.

{\bf 2) Other alkali lines and extra information:}

{\bf LOri036:} shows faint H$\alpha$ emission, but the strength of the alkali lines suggests that this object it older than the cluster members. Furthermore \citet{Sacco08} found a RV value not fully compatible with membership. For these two reasons we classify this object as possible non-member. 

{\bf LOri046, LOri049:} None of these sources shows H$\alpha$ emission; both have very low gravity according to the alkali absorption strengths; and both had indeed been classified already as possible non-members by \citet{Sacco08} because of their measured radial velocity. We classify as no members.

{\bf LOri110, LOri133, LOri141, LOri147, LOri151, LOri154, LOri165}: All these sources show surface gravities (based on the measured EW(NaI), see Table~\ref{tab:paramTOTAL}) too large compared with other cluster members. For the particular case of LOri110, no Li I was detected on our LRIS spectra, and \citet{Maxted08} also measured Na I EWs inconsistent with membership and confirmed that the radial velocities derived ruled out membership too. Regarding LOri133, LOri151 and LOri165, \citet{Barrado04} already classified these objects as doubtful candidate members based on their photometric properties

{\bf 3) Sources with spectral type earlier than M3 and no Lithium information:} 

As verified by \cite{Slesnick06a}, the Na I $\sim$8200 \AA~doublet strength saturates for spectral types earlier than M2. Furthermore, the differences among dwarfs, giants and young sources become very subtle already at M3 spectral types. Therefore, for these ``hotter" sources where we did not have spectra with resolution high enough to  measure Li, we have used different sets of criteria to assess membership:

{\bf - Non members: }

{\bf C69-IRAC-010 and C69-IRAC-008} with spectral types F9 and K3, show H$\alpha$ in absorption, no infrared excess at the IRAC wavelength range and a position in the HR diagram well below the 100 Myr isochrone, therefore we discard them as members.

{\bf - Members:}

{\bf C69XE-009} (K2 spectral type) shows a very intense H$\alpha$ emission and the estimated position in the HR diagram is perfectly compatible with membership.

{\bf C69-IRAC-001 and C69-IRAC-002} (M1.5 and M3 spectral types, respectively): Both are class II sources with thick disks according to the IRAC photometry and their positions in the HR diagram are compatible with membership.

{\bf LOri024, LOri061, LOri048 and LOri062} (spectral types from M1.5 to M3): All of them have been confirmed by \cite{Sacco08} and/or \cite{Maxted08} via radial velocity (and in all the cases, their positions in the HR diagram are compatible with membership).

{\bf C69-IRAC-007 and C69-IRAC-005} (M2.5 and M3): Both sources are active accretors (see Paper II) with thick disks and positions in the HR diagram compatible with membership.

{\bf C69XE-072} (M3 spectral type): Detected in our X-ray survey \citep{Barrado11} and with an estimated position in the HR diagram compatible with membership.

\begin{figure}
\centering
\includegraphics[width=9.0cm]{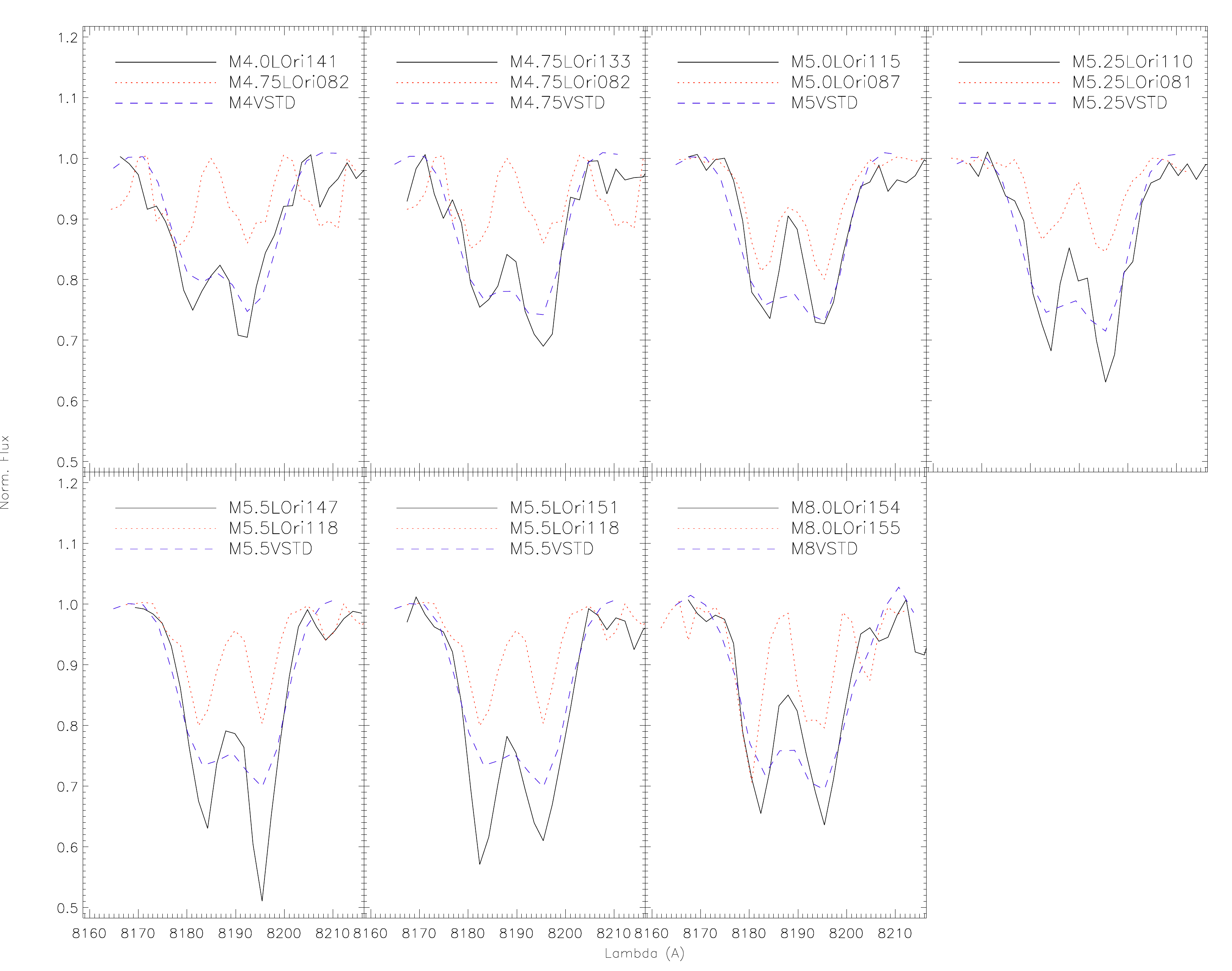}
\caption{Na I $\lambda$8200 doublet detail for some of the particular sources (all but LOri115 discarded as members) with CAFOS spectra. For each case we have over-plotted in red a confirmed member with the same spectral type, and in blue a field dwarf (again with the same spectral type and obtained in the same campaign). Note that for spectral types M4.75 and M5.25 we have averaged spectra of field dwarfs with spectral types M4.5 and M5.0, M5.0 and M5.5, respectively.
}
\label{fig:NaIno-memb}
\end{figure}

{\bf 4) Subaru L type sources:} The resolution of our Subaru/IRCS spectra was too low to resolve any alkali line. However, given the confirmed spectral type, their luminosities are not compatible with objects being much closer or much further than Lambda Orionis and therefore we assess Y? membership only expecting to confirm definitely membership by means of, for example, a proper motion study in a longer time baseline than the one provided by the images from which these candidates were selected (we only have one set of Subaru images; see \citealt{MoralesPhD})

To summarize and conclude the section, in Table~\ref{tab:paramTOTAL} we have included a column with the final membership assigned to each source. We consider members those labeled as Y?: objects showing some peculiarity already discussed but still considered as very good candidates for membership; and Y: spectroscopically confirmed members. In the latter case, we have included the reference of the work were the spectroscopic confirmation was achieved. 

There are $\sim$240 members from which 225 have been labeled as Y? or Y (36 of them are confirmed as members of Collinder 69 for the first time in this work). These confirmed members are displayed in an HR diagram in Fig~\ref{fig:HR}. We must note that there is a large dispersion on the diagram for the hotter part of the sample (T$_{\rm eff}$ larger than $\sim$3900 K), where almost 100\% of the candidates were confirmed by \cite{DM99, DM01}. However, if we take a closer look to the M (and cooler) population of Collinder 69 (where this work is focused); the best fitting isochrone is that of 5 Myr with an upper limit value of 20 Myr recovering 95\% of the confirmed members. The dispersion in this narrower area of the diagram, although less significant than the one affecting hotter sources, is mainly explained by objects harboring disks and has been previously addressed in \cite{Bayo08}.

Focusing only on the spectroscopically confirmed members (from this work and the literature), we find a brown dwarf to star ratio ($R_{\rm SS}$ as defined by \citealt{Briceno02}) of 0.06. This value is similar to that found for Taurus by \cite{Briceno02} but significantly smaller than the revised one by \cite{Guieu06} or the value reported for the ONC \citep{Kroupa03}. This fact can either pointing towards environmental effects in the substellar formation efficiency (as suggested previously in, for example \citealt{Briceno02,Kroupa03}) or towards pure limitations in the spatial extension of our surveys.

\section{Mass segregation: ejection mechanism.}
\label{sec:spatialdist}

As mentioned in Sec.~\ref{sec:intro}, one of the main goals of our long term project on the LOSFR is to investigate which of the currently proposed mechanisms of formation of brown dwarfs agrees best with the observations. 

According to the ejection scenario by \cite{Reipurth01}, brown dwarfs of an association such as Collider 69 should be located mostly around the exterior parts of the cluster. If we assume an escape velocity $\sim$1 km/s (as for the case of Taurus, \citealt{Kroupa03}), after $\sim$5 Myr, the ejected brown dwarfs should still have been recovered by some of our photometric surveys (a $\sim$35' distance is well covered by our SUBARU field and marginally by the Spitzer/IRAC one). 

\begin{figure*}[htbp]
\begin{center}
\includegraphics[width=9.cm,clip]{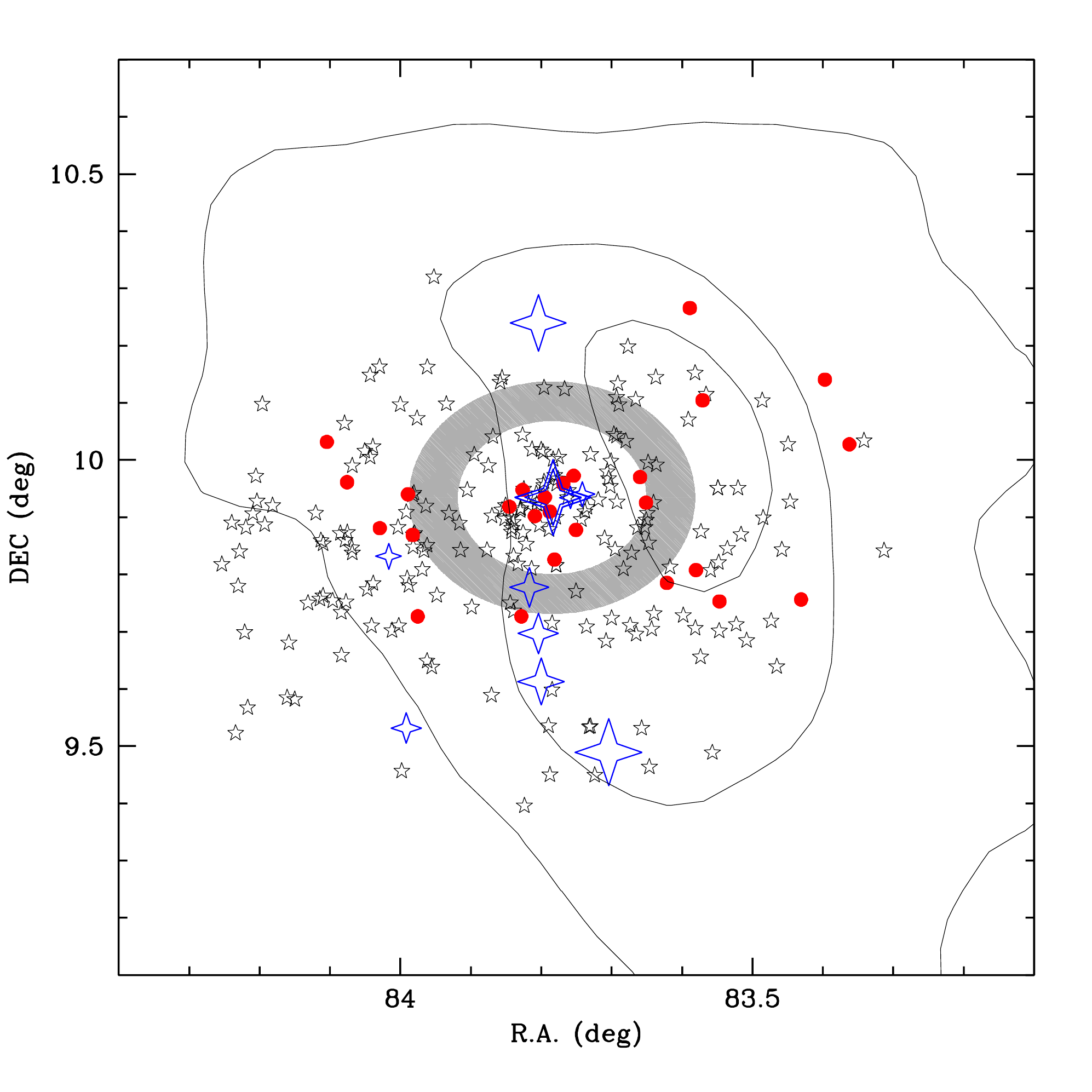}
\includegraphics[width=9.cm,clip]{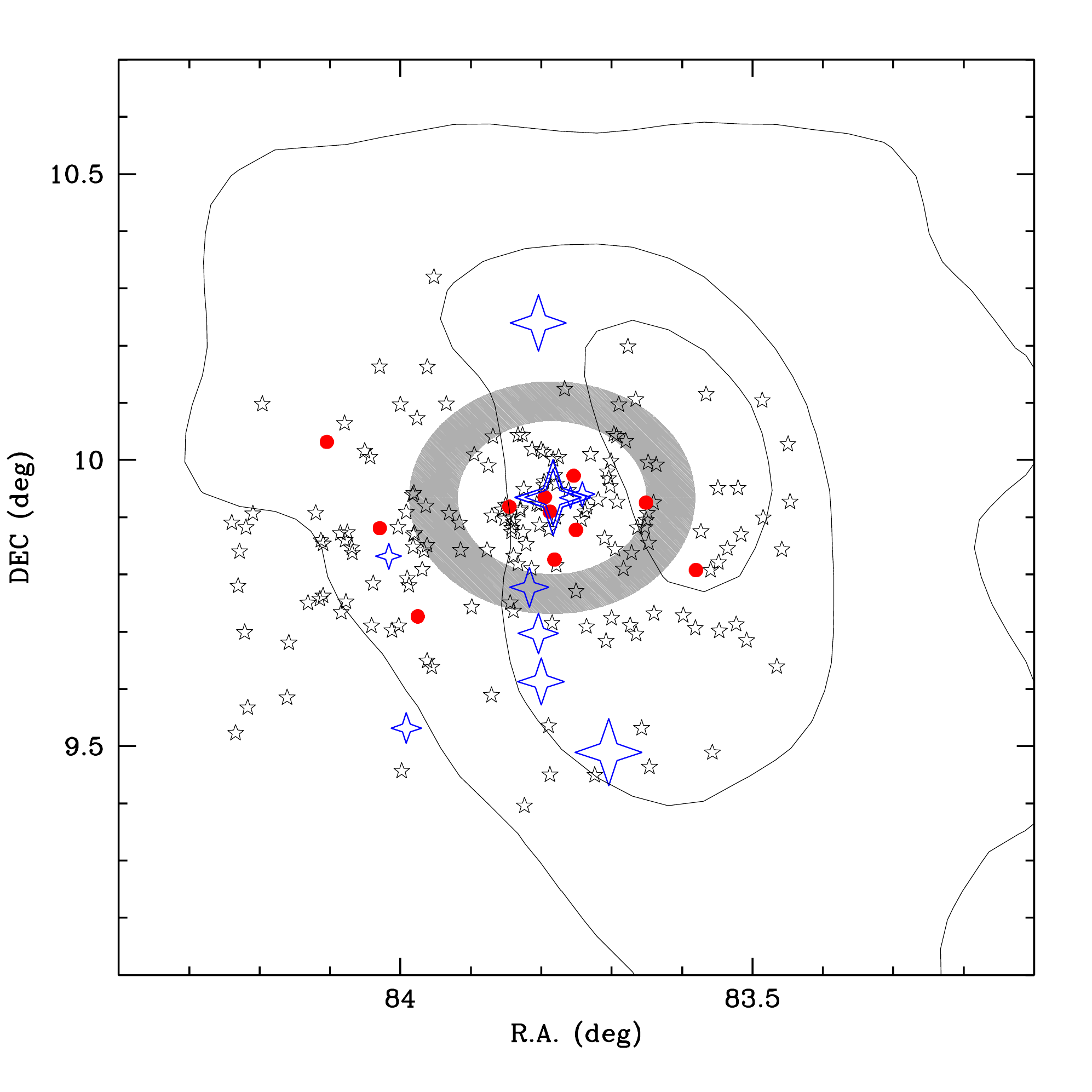}
\caption{Spatial distribution of: {\bf left panel:} good photometric candidates and {\bf right panel:} spectroscopically confirmed members. O and B populations are highlighted with large blue (four pointed) stars, members with masses above the H burning limit are displayed as small black (five pointed) stars and brown dwarfs as filled red dots. In both figures it is clear that brown dwarfs are not located at the edges of the cluster (as the ejection mechanism would predict). We have shaded in grey the ring corresponding to the void in the radial profile displayed on Fig.~\ref{fig:radial_dist}.}
\label{fig:dist_Esp}
\end{center}
\end{figure*}

To test this hypothesis, we have studied the spatial distribution of both stars and brown dwarfs in Collinder 69. In Fig.~\ref{fig:dist_Esp}, we plot two diagrams corresponding to the distribution of stellar and substellar confirmed and candidate members to Collinder 69 (left panel for the whole set of confirmed members and good photometric candidates, and the right one only showing the spectroscopically confirmed members). In both panels stellar sources are displayed as small black stars whilst substellar objects are highlighted with red dots. Even though what we can see in this figure is nothing but a projection, any structured grouping of the substellar population is far from obvious. 

From a more quantitative point of view, we tried to compare the distribution of stars and brown dwarfs with a two dimensional KS test, but because of the low numbers we are dealing with, the results were inconclusive.  Trying a more simplistic approach, since we are mainly interested in the radial distribution of brown dwarfs, we built the cumulative radial distribution of stellar and substellar sources (good candidate members and spectroscopically confirmed members, as in Fig.~\ref{fig:dist_Esp}). 

Although simple, this analysis had to be done carefully taking into account again the low numbers we are dealing with. To obtain the cumulative distributions in a robust way, we performed $\sim$60 different histograms using both multi-binning and multi-starting-point approaches and then smoothed the resulting function to make sure that the features found are bin and starting point independent.

\begin{figure}[htbp]
\begin{center}
\includegraphics[width=9.4cm,clip]{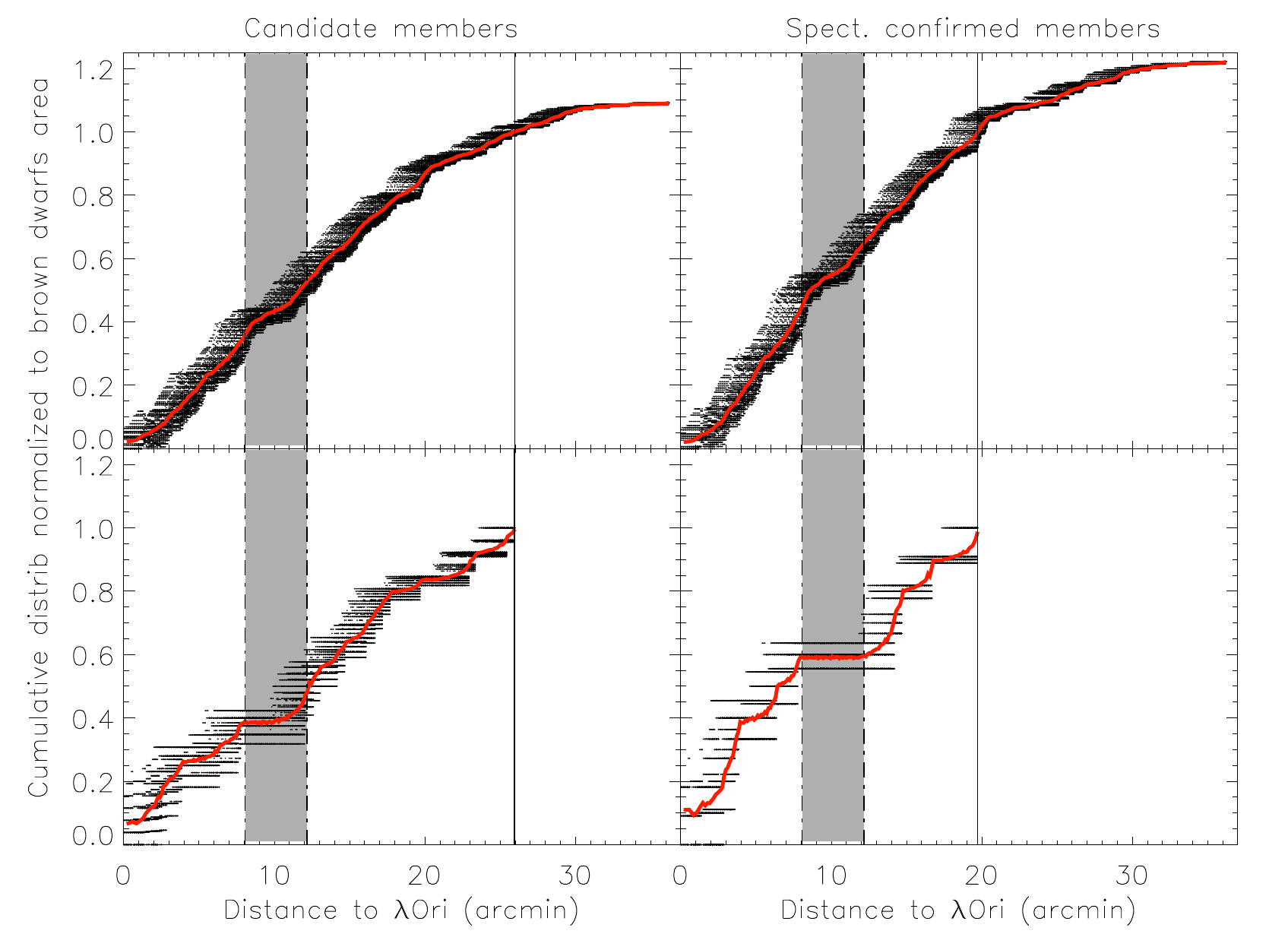}
\caption{Radial distribution of stars and brown dwarfs in Collinder 69: distance to $\lambda$ Ori vs. cumulative distribution of sources normalized to the radius where no more brown dwarfs are found in Collinder 69. As in the previous figure we include two sets of panels: on the left side candidates and confirmed members (top panel for stars and bottom for substellar objects) and on the right panel only spectroscopically confirmed members (as in the previous case, top for stars and bottom for substellar sources). Note the obvious void (grey shaded area) of brown dwarfs (and not so obvious but still present for stars too) from $\sim$8-12 arcmin distance to $\lambda$ Ori.}
\label{fig:radial_dist}
\end{center}
\end{figure}

We show the results of this method in Fig.~\ref{fig:radial_dist}: where the data-points of the individual histograms are drawn as black dots, and the smoothed final function is highlighted in red. It seems quite clear that there is no obvious differences in the distribution of stellar members of Collinder 69 (supposedly formed according to the classical paradigm)  and the substellar ones. Furthermore; if we compare the distribution of brown dwarfs with that of objects that are clearly stars (masses larger than 0.5 M$_{\odot}$), the results remain the same.

The only evident feature in these radial distributions is a void of both stars and brown dwarfs at $\sim$8-12 arcmin distance from $\lambda$ Ori (projected $\sim$1--1.4 pc). We have highlighted this void in Fig.~\ref{fig:dist_Esp} were we see no correlation of this area with differences in dust densities as traced by the IRAS contours (the solid black lines).

To try to quantify the physical relevance of this void, we have produced 100 synthetic 2D normal populations of sources. For each case we have produced three different sets of data: stars, brown dwarfs, and star + brown dwarfs. To draw each set, we have used the same number of objects that we have spectroscopically confirmed for each category in Collinder 69. The area from which the objects are drawn is the same area covered by our photometric surveys.

We have run a one dimensional KS test on the observed radial distribution compared to the synthetic ones (given the low number of brown dwarfs we have worked exclusively with the ``only stars" and ``stars + brown dwarfs" sets) and the result is that the probability of the observed radial distribution being drawn from a normal 2D population is negligible. Furthermore, only 7\% of the synthetic radial distributions show voids similar in width (but closer to the edges of the cluster) to the one present in our observed distribution. Therefore we are confident that this void is not just a statistical artifact.

Another interesting issue also not supporting the ejection scenario is the fact that Collinder 69 seems more extended in the stellar population than in the substellar one (the vertical solid line in Fig.~\ref{fig:radial_dist} marks the distance after which no more substellar members are found). This fact would indicate that the cluster has begun to be dynamically relaxed, so low-mass members are falling to the gravitational well. An alternative, given the very young age of the cluster, would be that the brown dwarfs might just have formed preferably closer to the center. As a note regarding the possible effect of the spatial coverage of our surveys, we want to point out that even though our SUBARU field reaches further separations from Lambda Orionis than the other surveys, we do not find brown dwarf candidates more distant than $\sim$20 arcmin.

In any case, we must remember once again that we are dealing with small numbers and that our photometric surveys are not wide enough to recover the possible ejected brown dwarfs if their velocity escape would be twice the one estimated for Taurus.  Therefore, the most we can conclude from our analysis is that the ejection mechanism of formation does not seem to be able to explain the observed properties of Collinder 69 for velocity escapes of $\sim$1 km/s or lower.

\section{The Initial Mass Function for spectroscopically confirmed members}
\label{sec:IMF}

\begin{figure}[htbp]
\begin{center}
\includegraphics[width=9.cm,clip]{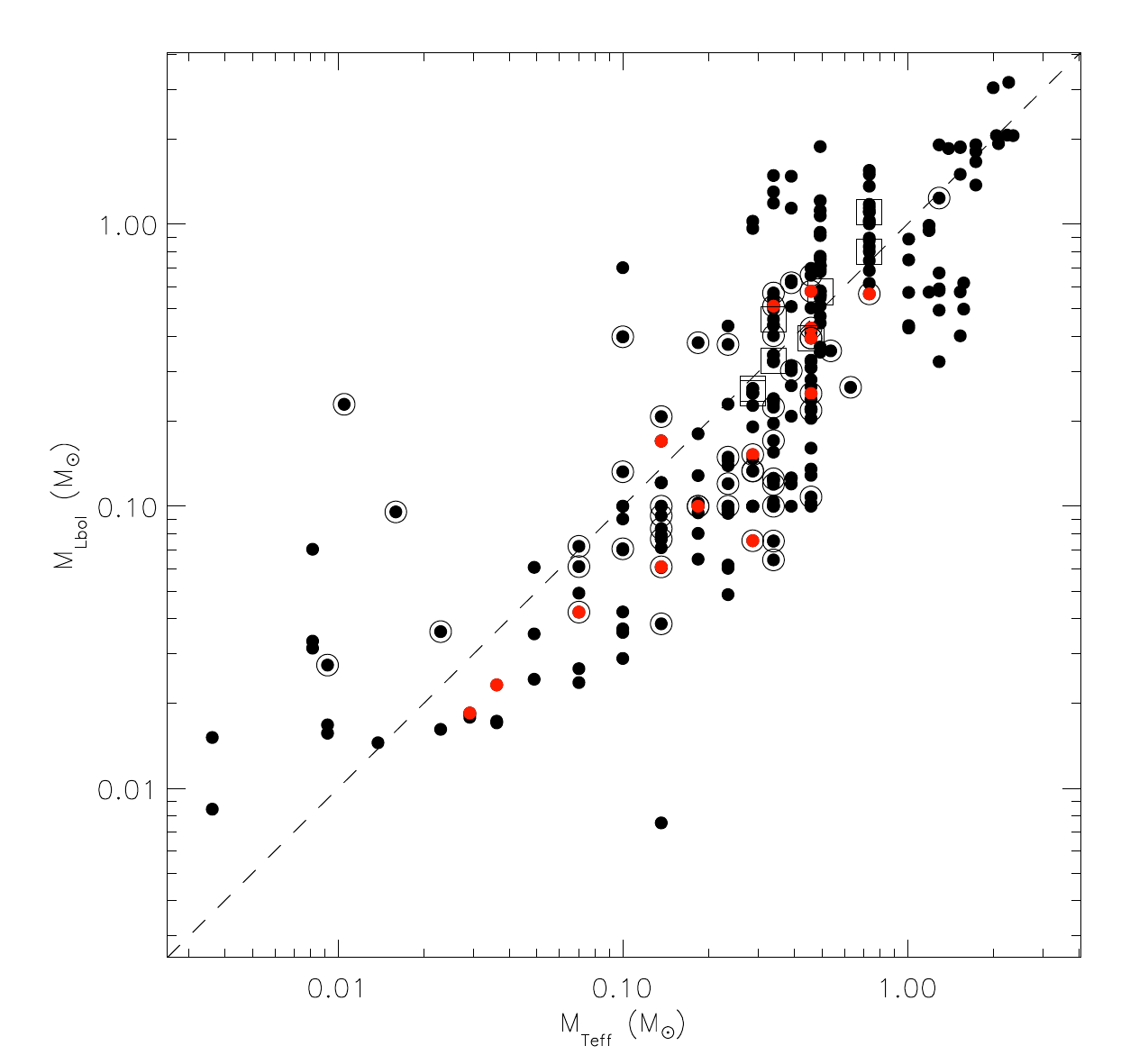}
\caption{Relationship between the masses derived from T$_{\rm eff}$ and L$_{\rm bol}$. Members (spectroscopically confirmed) and candidates (photometric) are displayed as solid dots. As in previous figures we have highlighted peculiar sources (Class II, binaries and accretors).}
\label{fig:MassTvsMassL}
\end{center}
\end{figure}

To derive the mass of the members of Collinder 69, we have used the estimated T$_{\rm eff}$ and L$_{\rm bol}$ from the SED fit along with a composite 5 Myr isochrone (\citealt{Siess00} + COND from \citealt{Baraffe03}). 
The average mass resulting from these two estimates for each object is provided in Table~\ref{tab:paramTOTAL}. 
As can be seen in Fig.~\ref{fig:MassTvsMassL} there is a quite large dispersion when comparing the estimates. If we separate our members according to the ratio between the two estimated masses into ``large dispersion" (those where one of the masses is larger than 1.5 times the other one) and ``acceptable'' dispersion (the ratio between the largest and the lowest mass lower than 1.5); we find that $\sim$30\% of the so-called ``large dispersion'' sources have been classified as Class II according to their IRAC colours, while the same class in the acceptable dispersion sample decreases to 15\% (this is not a question of scale since both samples have a similar number of sources). In fact, $\sim$65\% of the sources undergoing active accretion fall into the large dispersion sample. Therefore, our mass estimates are quite sensitive to the presence of disks (a caveat that does not surprise us, as has already been discussed in \citealt{Bayo08}). To take into account this uncertainty, in this section we will present all the calculations for both estimates of the mass of the members and for the average masses.

\begin{figure}[htbp]
\begin{center}
\includegraphics[width=9.cm,clip]{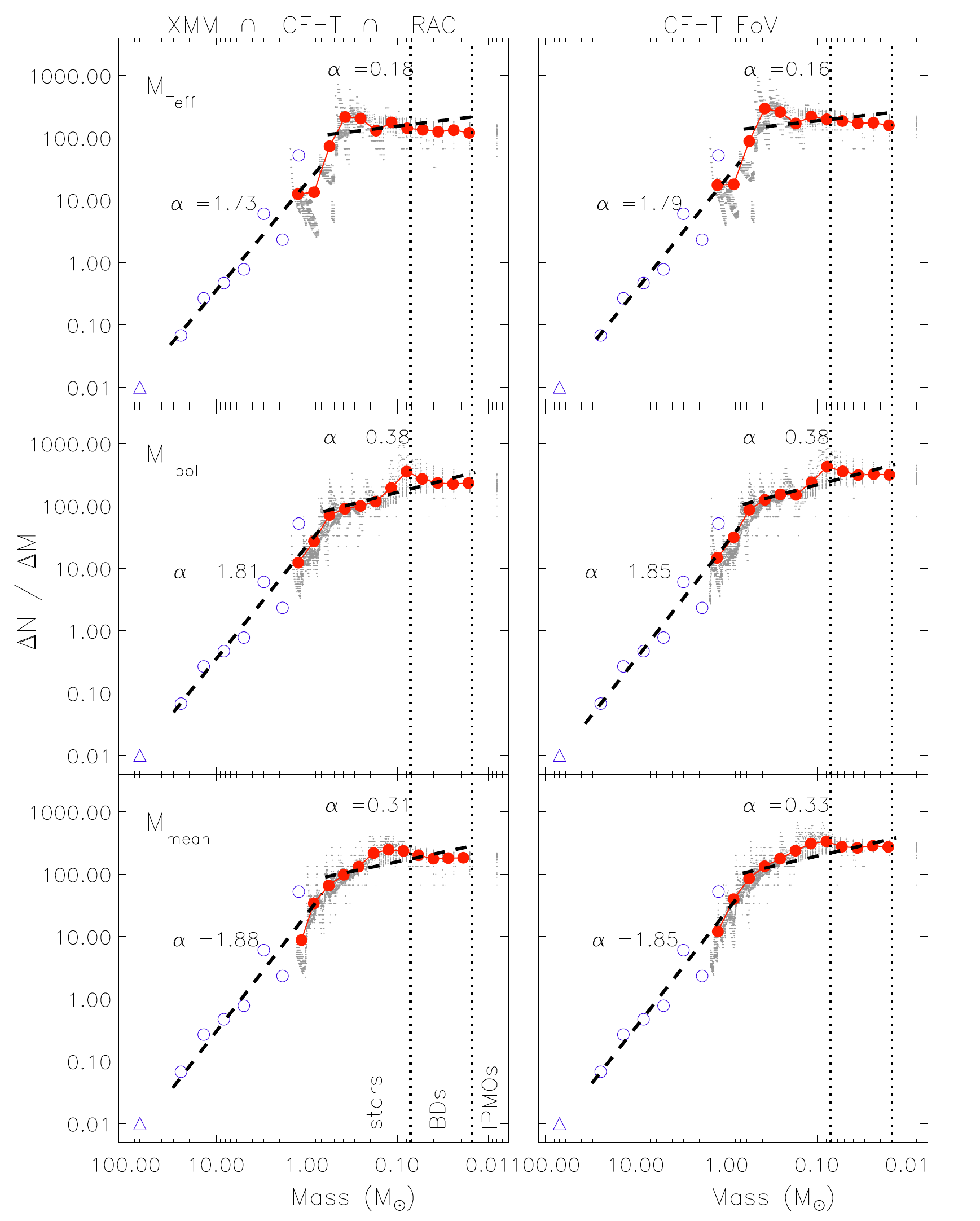}
\caption{IMFs derived for Collinder 69 taking into account three different determination of the mass for each members and two different areas within the cluster. {\bf Left panels:} Area confined within the intersection of XMM, CFHT and IRAC fields of view. {\bf Top:} IMF (I). Masses derived combining the estimated T$_{\rm eff}$ and the composite 5 Myr isochrone (SIESS + COND). {\bf Middle:} IMF (II). Masses derived combining the estimated L$_{\rm bol}$ and the composite 5 Myr isochrone (SIESS + COND). {\bf Bottom:} IMF (III). Masses derived averaging the masses from the previous two figures. {\bf Right panels:} Area confined within the CFHT field of view (including the central part of the cluster). {\bf Top:} IMF (IV). Same method to estimate masses as in IMF (I). {\bf Middle:} IMF(V). Same method to estimate masses as in IMF (II). {\bf Bottom:} IMF (VI). Same method to estimate masses as in IMF (III). }
\label{fig:IMFs}
\end{center}
\end{figure}

In the left panels of Fig.~\ref{fig:IMFs} we present the derived Initial Mass Functions (according to the different procedures followed to derive the masses of the sources) for Collinder 69. For this first case, we have considered only spectroscopically confirmed members confined within the intersection of the CFHT, Spitzer and XMM fields of view (see Fig.~\ref{fig:FoVs}). For the assembly of each IMF, in the range of masses covered by this work, we have proceeded in a similar manner as in the previous section with the multi-binning, multi-starting-point approach. Also, to enlarge the mass range coverage, we have included data from \cite{Murdin77} (scaling to the respective areas) as open blue circles. 
The last point at the massive end correspond to a possible SN (open blue triangle); see discussion on \cite{DM99}.

We must note that the previously described area (the intersection of the three FoVs), leaves the central part of the cluster (where the O8 III binary star, $\lambda$ Ori, is located) aside. In order to study the possible effects of this fact, 
in the right panels of Fig.~\ref{fig:IMFs} we show another set of IMFs (again three for the different estimated masses), this time focused on the CFHT field of view (scaling the XMM members to the corresponding area).

Regardless of the parameter used to infer the mass of our confirmed and candidate members, for masses above 0.65 M$_{\odot}$, the power law index of the IMF is similar to Salpeter's value, being much smaller for lower masses. In particular, the fitted slopes in the mass range 25--0.70 M$_{\odot}$ vary from 1.73 to 1.89; and those for the low and very low mass range, 0.70-0.01 M$_{\odot}$, correspond to much lower values; from 0.16 to 0.38 (see Table~\ref{tab:IMFslopes} for a summary of the slopes).

\begin{table}
\caption{Fitted slopes for the IMF of Collinder 69 taking into account the different methods to estimate the mass of the confirmed members (see text for further details). We also include the slopes estimated for low mass stars (the mass ranges are not always the same ones but roughly M$_{*} < $ 0.8 M$_{\odot}$) in other works for different open clusters.}
\label{tab:IMFslopes}
\begin{small}
\begin{tabular}{l@{\extracolsep{45pt}}c@{\extracolsep{-10pt}}cc}
\hline\hline
Estimated $\alpha$ for & & &\\
the mass range: & 25--0.65 M$_{\odot}$ && 0.65-0.01 M$_{\odot}$\\
\hline
C69 IMF (I) & 1.73 && 0.18\\
C69 IMF (II)  & 1.81 && 0.38\\
C69 IMF (III)  & 1.88 && 0.31\\
C69 IMF (IV)  & 1.79 && 0.16\\
C69 IMF (V)  & 1.85 && 0.38\\
C69 IMF (VI)  & 1.85 && 0.33\\
\hline
SOri$^1$ && 0.8&\\
APer$^{2,3}$ && 0.6&\\
M35$^4$ && (0.81) -- (-0.88)&\\
Pleiades$^5$ && 0.6&\\
\hline\hline
\end{tabular}
\end{small}
\begin{footnotesize}
\vspace{0.2cm} \\
Slope estimated with data from:\\
$^1$ \cite{Bejar01}\\
$^2$ \cite{Barrado01}\\
$^3$ \cite{Barrado02}\\
$^4$ \cite{Barrado01a}\\
$^5$ \cite{Bouvier98}\\
\end{footnotesize}
\end{table}

As we have mentioned before, the two areas for which we have derived the IMFs (the CFHT survey coverage and the intersection of CFHT, Spitzer and XMM) comprise populations that show distinct trends in terms of spatial distribution (the XMM coverage leaves a side a relatively large population of Class II sources). Surprisingly, this fact does not seem to affect the shape of the IMF since the slopes shown in Fig.~\ref{fig:IMFs} agree among themselves.

Our estimated slopes are systematically lower (but in a similar range) than those derived for photometrically compiled IMFs for several open clusters: SOri, APer, M35, and the Pleiades (see Table~\ref{tab:IMFslopes} for the exact numbers and references). For the low mass stars/substellar domain, the differences are not too large, regardless of the total mass, the environments and the age of each association. The only exception to this is the case of M35 --which corresponds to the core of the cluster--, a very rich association, about 150 Myr old, where important dynamical evolution might have taken place. 

\begin{figure}[htbp]
\begin{center}
\includegraphics[width=9.cm,clip]{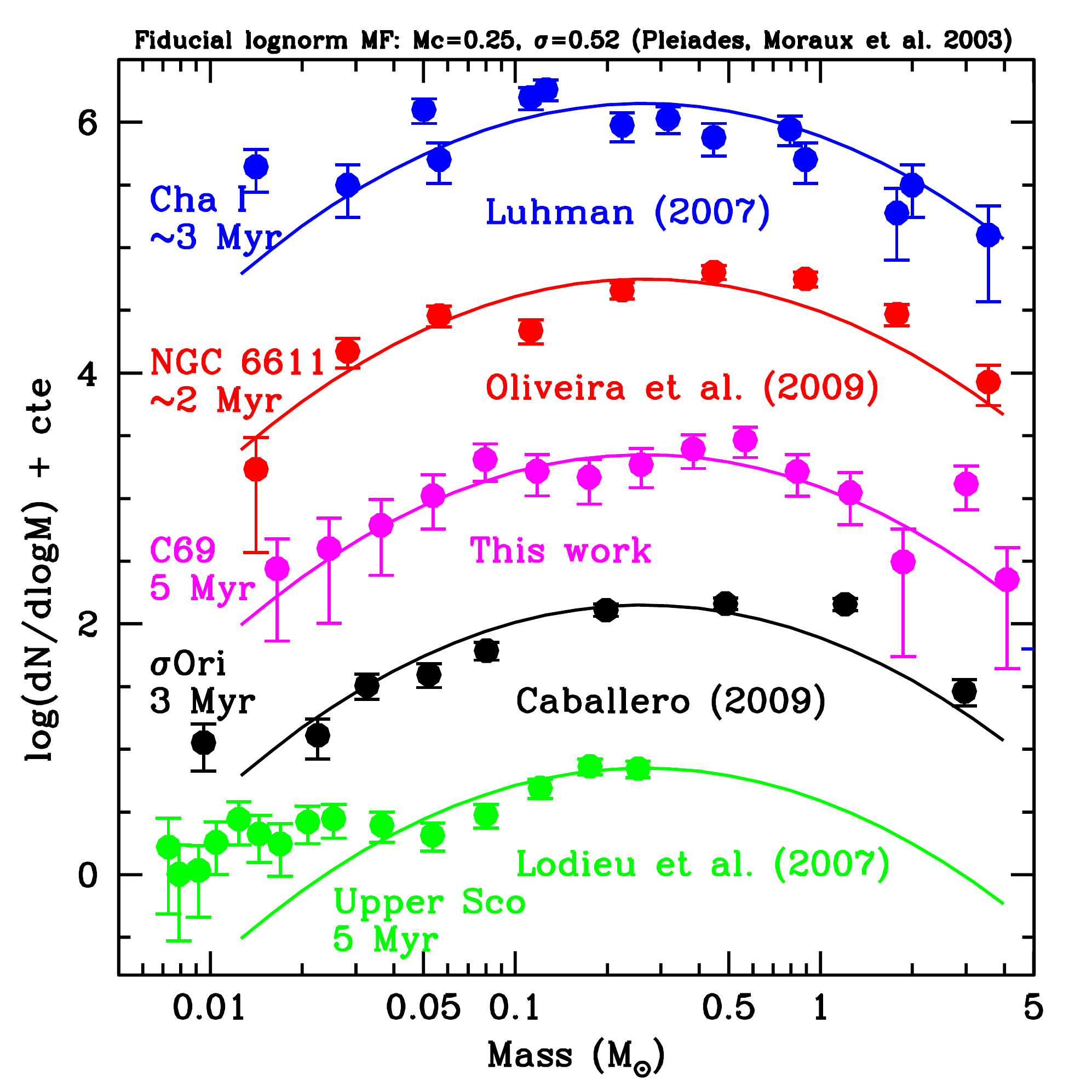}
\caption{Log-log representation of the IMF for different young associations. We compare our very complete spectroscopically confirmed IMF (in pink) with photometric ones derived for associations with similar ages ($\sim$2-5 Myr, \citealt{Luhman07,Oliveira09,Caballero09,Lodieu07}). The compilation of the data corresponding to clusters other than Collinder 69 are courtesy of J. Bouvier and E. Moraux.}
\label{fig:IMFsAssoc}
\end{center}
\end{figure}

In particular, if we only compare with similar age clusters, as we do in Fig.~\ref{fig:IMFsAssoc}, we can see that our IMF (in this case we have used the mean mass determination for the intersection of the three fields of view) looks very similar to that of, for example, NGC6611 with, in our case, a cut-off at $\sim$0.016M$_{\odot}$, were we think (because of the completeness of our photometric surveys) that we have reached the minimum mass for members of Collinder 69.

\section{Conclusions}
\label{sec:conclusions}

We have analyzed $\sim$170 optical and near infrared spectra with a wide range of resolutions of candidate members to the Collinder 69 cluster. Based on different criteria regarding molecular absorption bands, we have provided spectral types for all the sources. Using alkali lines as youth indicators, we have confirmed 90 members (and 9 possible members). For those sources of our survey overlapping with radial velocity surveys (by \citealt{Sacco08, Maxted08}), we obtain very similar results regarding membership.
A summary of the sources analyzed in this work can be found in Table~\ref{tab:paramTOTAL}.

We have confirmed the cool nature of {\bf the lowest mass candidate members to Collinder 69} reported so far. With derived spectral types L0-L2 corresponding to effective temperatures of $\sim$ 2000 K (according to the spectroscopic temperature scales derived by \citealt{Basri00}), if members, they have estimated masses, following the 5 Myr DUSTY isochrone by \citealt{Chabrier00}, as low as 0.016 M$_{\odot}$ ($\sim$16.5 M$_{\rm Jupiter}$).

We have built the most complete IMF based on spectroscopically confirmed members for a young association and when compared with others (photometric IMFs), no significant differences in the low mass and substellar domain are identified.

With our study of the spatial distribution of confirmed stellar and substellar members of Collinder 69, we provide some caveats to the ejection formation scenario for brown dwarfs, in order for it to be feasible in this region.

\begin{acknowledgements}
We want to thank the anonymous referee for very useful comments that improved the manuscript with no doubt.
This research has been funded by Spanish grants ESP2007-65475-C02-02, CSD2006-00070, PRICIT-S2009/ESP-1496 and AYA 2010-21161-C02-02. The support of the ESAC faculty is also recognized. It makes use of VOSA, developed under the Spanish Virtual Observatory project supported from the Spanish MICINN through grant AyA2008-02156, and of the SIMBAD database, operated at CDS, Strasbourg, France.
The authors wish to thank J. Bouvier and E. Moraux for their data compilation on IMFs of different young associations and Y. Momany for the interesting discussion about statistics when building histograms.
This research has made use of the Keck Observatory Archive (KOA), which is operated by the W. M. Keck Observatory and the NASA Exoplanet Science Institute (NExScI), under contract with the National Aeronautics and Space Administration. 
This paper includes data gathered with the 6.5 meter Magellan Telescopes located at Las Campanas Observatory, Chile.
Based on observations made with ESO Telescopes at the La Silla Paranal Observatory under programmes ID 080.C-0592 and  078.C-0124. Based on observations collected at the Centro Astron\'omico Hispano Alem\'an (CAHA) at Calar Alto, operated jointly by the Max-Planck Institut f\"ur Astronomie and the Instituto de Astrof\'isica de Andaluc\'ia (CSIC).
Based in part on data collected at Subaru Telescope, which is operated by the National Astronomical Observatory of Japan.
\end{acknowledgements}

\bibliographystyle{aa}
\bibliography{./biblio}

\begin{thebibliography}{81}
\expandafter\ifx\csname natexlab\endcsname\relax\def\natexlab#1{#1}\fi

\bibitem[{{Allard} {et~al.}(2003){Allard}, {Guillot}, {Ludwig}, {Hauschildt},
  {Schweitzer}, {Alexander}, \& {Ferguson}}]{Allard03}
{Allard}, F., {Guillot}, T., {Ludwig}, H.-G., {et~al.} 2003, in IAU Symposium,
  Vol. 211, Brown Dwarfs, ed. E.~{Mart{\'{\i}}n}, 325--+

\bibitem[{{Allard} {et~al.}(2001){Allard}, {Hauschildt}, {Alexander},
  {Tamanai}, \& {Schweitzer}}]{Allard01}
{Allard}, F., {Hauschildt}, P.~H., {Alexander}, D.~R., {Tamanai}, A., \&
  {Schweitzer}, A. 2001, \apj, 556, 357

\bibitem[{{Baraffe} \& {Chabrier}(2010)}]{Baraffe10}
{Baraffe}, I. \& {Chabrier}, G. 2010, \aap, 521, A44+

\bibitem[{{Baraffe} {et~al.}(1998){Baraffe}, {Chabrier}, {Allard}, \&
  {Hauschildt}}]{Baraffe98}
{Baraffe}, I., {Chabrier}, G., {Allard}, F., \& {Hauschildt}, P. 1998, VizieR
  Online Data Catalog, 333, 70403

\bibitem[{{Baraffe} {et~al.}(2002){Baraffe}, {Chabrier}, {Allard}, \&
  {Hauschildt}}]{Baraffe02}
{Baraffe}, I., {Chabrier}, G., {Allard}, F., \& {Hauschildt}, P.~H. 2002, \aap,
  382, 563

\bibitem[{{Baraffe} {et~al.}(2003){Baraffe}, {Chabrier}, {Barman}, {Allard}, \&
  {Hauschildt}}]{Baraffe03}
{Baraffe}, I., {Chabrier}, G., {Barman}, T.~S., {Allard}, F., \& {Hauschildt},
  P.~H. 2003, \aap, 402, 701

\bibitem[{{Barrado} {et~al.}(2011){Barrado}, {Stelzer}, {Morales-Calder{\'o}n},
  {Bayo}, {Hu{\'e}lamo}, {Stauffer}, {Hodgkin}, {Galindo}, \&
  {Verdugo}}]{Barrado11}
{Barrado}, D., {Stelzer}, B., {Morales-Calder{\'o}n}, M., {et~al.} 2011, \aap,
  526, A21+

\bibitem[{{Barrado y Navascu{\'e}s} {et~al.}(2001{\natexlab{a}}){Barrado y
  Navascu{\'e}s}, {Garc{\'{\i}}a L{\'o}pez}, {Severino}, \&
  {Gomez}}]{Barrado01}
{Barrado y Navascu{\'e}s}, D., {Garc{\'{\i}}a L{\'o}pez}, R.~J., {Severino},
  G., \& {Gomez}, M.~T. 2001{\natexlab{a}}, \aap, 371, 652

\bibitem[{{Barrado y Navascu{\'e}s} \& {Mart{\'{\i}}n}(2003)}]{Barrado03}
{Barrado y Navascu{\'e}s}, D. \& {Mart{\'{\i}}n}, E.~L. 2003, \aj, 126, 2997

\bibitem[{{Barrado y Navascu{\'e}s} {et~al.}(2004{\natexlab{a}}){Barrado y
  Navascu{\'e}s}, {Mohanty}, \& {Jayawardhana}}]{Barrado04a}
{Barrado y Navascu{\'e}s}, D., {Mohanty}, S., \& {Jayawardhana}, R.
  2004{\natexlab{a}}, \apj, 604, 284

\bibitem[{{Barrado y Navascu{\'e}s} {et~al.}(2004{\natexlab{b}}){Barrado y
  Navascu{\'e}s}, {Stauffer}, {Bouvier}, {Jayawardhana}, \&
  {Cuillandre}}]{Barrado04b}
{Barrado y Navascu{\'e}s}, D., {Stauffer}, J.~R., {Bouvier}, J.,
  {Jayawardhana}, R., \& {Cuillandre}, J.-C. 2004{\natexlab{b}}, \apj, 610,
  1064

\bibitem[{{Barrado y Navascu{\'e}s} {et~al.}(2004{\natexlab{c}}){Barrado y
  Navascu{\'e}s}, {Stauffer}, {Bouvier}, {Jayawardhana}, \&
  {Cuillandre}}]{Barrado04}
{Barrado y Navascu{\'e}s}, D., {Stauffer}, J.~R., {Bouvier}, J.,
  {Jayawardhana}, R., \& {Cuillandre}, J.-C. 2004{\natexlab{c}}, \apj, 610,
  1064

\bibitem[{{Barrado y Navascu{\'e}s} {et~al.}(2001{\natexlab{b}}){Barrado y
  Navascu{\'e}s}, {Stauffer}, {Bouvier}, \& {Mart{\'{\i}}n}}]{Barrado01a}
{Barrado y Navascu{\'e}s}, D., {Stauffer}, J.~R., {Bouvier}, J., \&
  {Mart{\'{\i}}n}, E.~L. 2001{\natexlab{b}}, \apj, 546, 1006

\bibitem[{{Barrado y Navascu{\'e}s} {et~al.}(2007){Barrado y Navascu{\'e}s},
  {Stauffer}, {Morales-Calder{\'o}n}, {Bayo}, {Fazzio}, {Megeath}, {Allen},
  {Hartmann}, \& {Calvet}}]{Barrado07}
{Barrado y Navascu{\'e}s}, D., {Stauffer}, J.~R., {Morales-Calder{\'o}n}, M.,
  {et~al.} 2007, \apj, 664, 481

\bibitem[{{Barrado y Navascu{\'e}s} {et~al.}(1999){Barrado y Navascu{\'e}s},
  {Stauffer}, \& {Patten}}]{Barrado99}
{Barrado y Navascu{\'e}s}, D., {Stauffer}, J.~R., \& {Patten}, B.~M. 1999,
  \apjl, 522, L53

\bibitem[{{Barrado y Navascu{\'e}s} {et~al.}(2002){Barrado y Navascu{\'e}s},
  {Zapatero Osorio}, {Mart{\'{\i}}n}, {B{\'e}jar}, {Rebolo}, \&
  {Mundt}}]{Barrado02}
{Barrado y Navascu{\'e}s}, D., {Zapatero Osorio}, M.~R., {Mart{\'{\i}}n},
  E.~L., {et~al.} 2002, \aap, 393, L85

\bibitem[{{Basri}(1997)}]{Basri97}
{Basri}, G. 1997, Memorie della Societa Astronomica Italiana, 68, 917

\bibitem[{{Basri} {et~al.}(2000){Basri}, {Mohanty}, {Allard}, {Hauschildt},
  {Delfosse}, {Mart{\'{\i}}n}, {Forveille}, \& {Goldman}}]{Basri00}
{Basri}, G., {Mohanty}, S., {Allard}, F., {et~al.} 2000, \apj, 538, 363

\bibitem[{{Bayo}(2009)}]{BayoPhD}
{Bayo}, A. 2009, PhD thesis, UNIVERSIDAD AUT\'ONOMA DE MADRID

\bibitem[{{Bayo} {et~al.}(2008){Bayo}, {Rodrigo}, {Barrado Y Navascu{\'e}s},
  {Solano}, {Guti{\'e}rrez}, {Morales-Calder{\'o}n}, \& {Allard}}]{Bayo08}
{Bayo}, A., {Rodrigo}, C., {Barrado Y Navascu{\'e}s}, D., {et~al.} 2008, \aap,
  492, 277

\bibitem[{{B{\'e}jar} {et~al.}(2001){B{\'e}jar}, {Mart{\'{\i}}n}, {Zapatero
  Osorio}, {Rebolo}, {Barrado y Navascu{\'e}s}, {Bailer-Jones}, {Mundt},
  {Baraffe}, {Chabrier}, \& {Allard}}]{Bejar01}
{B{\'e}jar}, V.~J.~S., {Mart{\'{\i}}n}, E.~L., {Zapatero Osorio}, M.~R.,
  {et~al.} 2001, \apj, 556, 830

\bibitem[{{Bessell} {et~al.}(1998){Bessell}, {Castelli}, \& {Plez}}]{Bessell98}
{Bessell}, M.~S., {Castelli}, F., \& {Plez}, B. 1998, \aap, 337, 321

\bibitem[{{Bildsten} {et~al.}(1997){Bildsten}, {Brown}, {Matzner}, \&
  {Ushomirsky}}]{Bildsten97}
{Bildsten}, L., {Brown}, E.~F., {Matzner}, C.~D., \& {Ushomirsky}, G. 1997,
  \apj, 482, 442

\bibitem[{{Bouvier} {et~al.}(1998){Bouvier}, {Stauffer}, {Martin}, {Barrado y
  Navascues}, {Wallace}, \& {Bejar}}]{Bouvier98}
{Bouvier}, J., {Stauffer}, J.~R., {Martin}, E.~L., {et~al.} 1998, \aap, 336,
  490

\bibitem[{{Brice{\~n}o} {et~al.}(2002){Brice{\~n}o}, {Luhman}, {Hartmann},
  {Stauffer}, \& {Kirkpatrick}}]{Briceno02}
{Brice{\~n}o}, C., {Luhman}, K.~L., {Hartmann}, L., {Stauffer}, J.~R., \&
  {Kirkpatrick}, J.~D. 2002, \apj, 580, 317

\bibitem[{{Burrows} {et~al.}(1997){Burrows}, {Marley}, {Hubbard}, {Lunine},
  {Guillot}, {Saumon}, {Freedman}, {Sudarsky}, \& {Sharp}}]{Burrows97}
{Burrows}, A., {Marley}, M., {Hubbard}, W.~B., {et~al.} 1997, \apj, 491, 856

\bibitem[{{Caballero} {et~al.}(2009){Caballero}, {L{\'o}pez-Santiago}, {de
  Castro}, \& {Cornide}}]{Caballero09}
{Caballero}, J.~A., {L{\'o}pez-Santiago}, J., {de Castro}, E., \& {Cornide}, M.
  2009, \aj, 137, 5012

\bibitem[{{Castelli} {et~al.}(1997){Castelli}, {Gratton}, \&
  {Kurucz}}]{Castelli97}
{Castelli}, F., {Gratton}, R.~G., \& {Kurucz}, R.~L. 1997, \aap, 318, 841

\bibitem[{{Chabrier} \& {Baraffe}(1997)}]{Chabrier97}
{Chabrier}, G. \& {Baraffe}, I. 1997, \aap, 327, 1039

\bibitem[{{Chabrier} {et~al.}(2000){Chabrier}, {Baraffe}, {Allard}, \&
  {Hauschildt}}]{Chabrier00}
{Chabrier}, G., {Baraffe}, I., {Allard}, F., \& {Hauschildt}, P. 2000, \apj,
  542, 464

\bibitem[{{Comer{\'o}n} {et~al.}(2000){Comer{\'o}n}, {Neuh{\"a}user}, \&
  {Kaas}}]{Comeron00}
{Comer{\'o}n}, F., {Neuh{\"a}user}, R., \& {Kaas}, A.~A. 2000, \aap, 359, 269

\bibitem[{{Cruz} \& {Reid}(2002)}]{Cruz02}
{Cruz}, K.~L. \& {Reid}, I.~N. 2002, \aj, 123, 2828

\bibitem[{{da Silva} {et~al.}(2009){da Silva}, {Torres}, {de La Reza}, {Quast},
  {Melo}, \& {Sterzik}}]{daSilva09}
{da Silva}, L., {Torres}, C.~A.~O., {de La Reza}, R., {et~al.} 2009, \aap, 508,
  833

\bibitem[{{Dolan} \& {Mathieu}(1999)}]{DM99}
{Dolan}, C.~J. \& {Mathieu}, R.~D. 1999, \aj, 118, 2409

\bibitem[{{Dolan} \& {Mathieu}(2001)}]{DM01}
{Dolan}, C.~J. \& {Mathieu}, R.~D. 2001, \aj, 121, 2124

\bibitem[{{Dolan} \& {Mathieu}(2002)}]{Dolan02}
{Dolan}, C.~J. \& {Mathieu}, R.~D. 2002, \aj, 123, 387

\bibitem[{{Duerr} {et~al.}(1982){Duerr}, {Imhoff}, \& {Lada}}]{Duerr82}
{Duerr}, R., {Imhoff}, C.~L., \& {Lada}, C.~J. 1982, \apj, 261, 135

\bibitem[{{Fern{\'a}ndez} \& {Comer{\'o}n}(2001)}]{Fernandez01}
{Fern{\'a}ndez}, M. \& {Comer{\'o}n}, F. 2001, \aap, 380, 264

\bibitem[{{Geballe} {et~al.}(2002){Geballe}, {Knapp}, {Leggett}, {Fan},
  {Golimowski}, {Anderson}, {Brinkmann}, {Csabai}, {Gunn}, {Hawley},
  {Hennessy}, {Henry}, {Hill}, {Hindsley}, {Ivezi{\'c}}, {Lupton}, {McDaniel},
  {Munn}, {Narayanan}, {Peng}, {Pier}, {Rockosi}, {Schneider}, {Smith},
  {Strauss}, {Tsvetanov}, {Uomoto}, {York}, \& {Zheng}}]{Geballe02}
{Geballe}, T.~R., {Knapp}, G.~R., {Leggett}, S.~K., {et~al.} 2002, \apj, 564,
  466

\bibitem[{{Gomez} \& {Lada}(1998)}]{Gomez98}
{Gomez}, M. \& {Lada}, C.~J. 1998, \aj, 115, 1524

\bibitem[{{G{\'o}mez} \& {Persi}(2002)}]{Gomez02}
{G{\'o}mez}, M. \& {Persi}, P. 2002, \aap, 389, 494

\bibitem[{{Guieu} {et~al.}(2006){Guieu}, {Dougados}, {Monin}, {Magnier}, \&
  {Mart{\'{\i}}n}}]{Guieu06}
{Guieu}, S., {Dougados}, C., {Monin}, J.-L., {Magnier}, E., \& {Mart{\'{\i}}n},
  E.~L. 2006, \aap, 446, 485

\bibitem[{{Hauschildt} {et~al.}(1999){Hauschildt}, {Allard}, \&
  {Baron}}]{Hauschildt99}
{Hauschildt}, P.~H., {Allard}, F., \& {Baron}, E. 1999, \apj, 512, 377

\bibitem[{{Jeffries} {et~al.}(2009){Jeffries}, {Jackson}, {James}, \&
  {Cargile}}]{Jeffries09}
{Jeffries}, R.~D., {Jackson}, R.~J., {James}, D.~J., \& {Cargile}, P.~A. 2009,
  \mnras, 400, 317

\bibitem[{{Jeffries} \& {Oliveira}(2005)}]{Jeffries05}
{Jeffries}, R.~D. \& {Oliveira}, J.~M. 2005, \mnras, 358, 13

\bibitem[{{Jeffries} {et~al.}(2003){Jeffries}, {Oliveira}, {Barrado y
  Navascu{\'e}s}, \& {Stauffer}}]{Jeffries03}
{Jeffries}, R.~D., {Oliveira}, J.~M., {Barrado y Navascu{\'e}s}, D., \&
  {Stauffer}, J.~R. 2003, \mnras, 343, 1271

\bibitem[{{Jones} {et~al.}(1994){Jones}, {Longmore}, {Jameson}, \&
  {Mountain}}]{Jones94}
{Jones}, H.~R.~A., {Longmore}, A.~J., {Jameson}, R.~F., \& {Mountain}, C.~M.
  1994, \mnras, 267, 413

\bibitem[{{Kroupa} \& {Bouvier}(2003)}]{Kroupa03}
{Kroupa}, P. \& {Bouvier}, J. 2003, \mnras, 346, 369

\bibitem[{{Leggett} {et~al.}(2000){Leggett}, {Allard}, {Dahn}, {Hauschildt},
  {Kerr}, \& {Rayner}}]{Leggett00}
{Leggett}, S.~K., {Allard}, F., {Dahn}, C., {et~al.} 2000, \apj, 535, 965

\bibitem[{{Lodieu} {et~al.}(2008){Lodieu}, {Hambly}, {Jameson}, \&
  {Hodgkin}}]{Lodieu08}
{Lodieu}, N., {Hambly}, N.~C., {Jameson}, R.~F., \& {Hodgkin}, S.~T. 2008,
  \mnras, 383, 1385

\bibitem[{{Lodieu} {et~al.}(2007){Lodieu}, {Hambly}, {Jameson}, {Hodgkin},
  {Carraro}, \& {Kendall}}]{Lodieu07}
{Lodieu}, N., {Hambly}, N.~C., {Jameson}, R.~F., {et~al.} 2007, \mnras, 374,
  372

\bibitem[{{Luhman}(1999)}]{Luhman99}
{Luhman}, K.~L. 1999, \apj, 525, 466

\bibitem[{{Luhman} {et~al.}(2007){Luhman}, {Adame}, {D'Alessio}, {Calvet},
  {McLeod}, {Bohac}, {Forrest}, {Hartmann}, {Sargent}, \& {Watson}}]{Luhman07}
{Luhman}, K.~L., {Adame}, L., {D'Alessio}, P., {et~al.} 2007, \apj, 666, 1219

\bibitem[{{Luhman} {et~al.}(1997){Luhman}, {Liebert}, \& {Rieke}}]{Luhman97}
{Luhman}, K.~L., {Liebert}, J., \& {Rieke}, G.~H. 1997, \apjl, 489, L165+

\bibitem[{{Luhman} {et~al.}(2003){Luhman}, {Stauffer}, {Muench}, {Rieke},
  {Lada}, {Bouvier}, \& {Lada}}]{Luhman03}
{Luhman}, K.~L., {Stauffer}, J.~R., {Muench}, A.~A., {et~al.} 2003, \apj, 593,
  1093

\bibitem[{{Mart{\'{\i}}n} {et~al.}(1999){Mart{\'{\i}}n}, {Delfosse}, {Basri},
  {Goldman}, {Forveille}, \& {Zapatero Osorio}}]{Martin99}
{Mart{\'{\i}}n}, E.~L., {Delfosse}, X., {Basri}, G., {et~al.} 1999, \aj, 118,
  2466

\bibitem[{{Martin} {et~al.}(1996){Martin}, {Rebolo}, \&
  {Zapatero-Osorio}}]{Martin96}
{Martin}, E.~L., {Rebolo}, R., \& {Zapatero-Osorio}, M.~R. 1996, \apj, 469, 706

\bibitem[{{Maxted} {et~al.}(2008){Maxted}, {Jeffries}, {Oliveira}, {Naylor}, \&
  {Jackson}}]{Maxted08}
{Maxted}, P.~F.~L., {Jeffries}, R.~D., {Oliveira}, J.~M., {Naylor}, T., \&
  {Jackson}, R.~J. 2008, \mnras, 385, 2210

\bibitem[{{Mohanty} {et~al.}(2005){Mohanty}, {Jayawardhana}, \&
  {Basri}}]{Mohanty05}
{Mohanty}, S., {Jayawardhana}, R., \& {Basri}, G. 2005, \apj, 626, 498

\bibitem[{{Morales-Calder{\'o}n}(2008)}]{MoralesPhD}
{Morales-Calder{\'o}n}, M. 2008, PhD thesis, UNIVERSIDAD AUT\'ONOMA DE MADRID

\bibitem[{Mouschovias(1991)}]{Mouschovias91}
Mouschovias, T.~C. 1991, NATO ASIC, Proc. 342, 449

\bibitem[{{Murdin} \& {Penston}(1977)}]{Murdin77}
{Murdin}, P. \& {Penston}, M.~V. 1977, \mnras, 181, 657

\bibitem[{{Natta} {et~al.}(2004){Natta}, {Testi}, {Muzerolle}, {Randich},
  {Comer{\'o}n}, \& {Persi}}]{Natta04}
{Natta}, A., {Testi}, L., {Muzerolle}, J., {et~al.} 2004, \aap, 424, 603

\bibitem[{{Neuhaeuser} {et~al.}(1998){Neuhaeuser}, {Wolk}, {Torres},
  {Preibisch}, {Stout-Batalha}, {Hatzes}, {Frink}, {Wichmann}, {Covino},
  {Alcala}, {Brandner}, {Walter}, {Sterzik}, \& {Koehler}}]{Neuhaeuser98}
{Neuhaeuser}, R., {Wolk}, S.~J., {Torres}, G., {et~al.} 1998, \aap, 334, 873

\bibitem[{{Oliveira} {et~al.}(2009){Oliveira}, {Jeffries}, \& {van
  Loon}}]{Oliveira09}
{Oliveira}, J.~M., {Jeffries}, R.~D., \& {van Loon}, J.~T. 2009, \mnras, 392,
  1034

\bibitem[{{Padoan} \& {Nordlund}(2002)}]{Padoan02}
{Padoan}, P. \& {Nordlund}, {\AA}. 2002, \apj, 576, 870

\bibitem[{{Reid} \& {Gizis}(1997)}]{Reid97}
{Reid}, I.~N. \& {Gizis}, J.~E. 1997, \aj, 113, 2246

\bibitem[{{Reid} {et~al.}(1995){Reid}, {Hawley}, \& {Gizis}}]{Reid95}
{Reid}, I.~N., {Hawley}, S.~L., \& {Gizis}, J.~E. 1995, \aj, 110, 1838

\bibitem[{{Reipurth} \& {Clarke}(2001)}]{Reipurth01}
{Reipurth}, B. \& {Clarke}, C. 2001, \aj, 122, 432

\bibitem[{{Sacco} {et~al.}(2008){Sacco}, {Franciosini}, {Randich}, \&
  {Pallavicini}}]{Sacco08}
{Sacco}, G.~G., {Franciosini}, E., {Randich}, S., \& {Pallavicini}, R. 2008,
  \aap, 488, 167

\bibitem[{{Schiavon} {et~al.}(1997){Schiavon}, {Barbuy}, {Rossi}, \&
  {Milone}}]{Schiavon97}
{Schiavon}, R.~P., {Barbuy}, B., {Rossi}, S.~C.~F., \& {Milone}, A. 1997, \apj,
  479, 902

\bibitem[{{Shu} {et~al.}(1987){Shu}, {Adams}, \& {Lizano}}]{Shu87}
{Shu}, F.~H., {Adams}, F.~C., \& {Lizano}, S. 1987, \araa, 25, 23

\bibitem[{{Siess} {et~al.}(2000){Siess}, {Dufour}, \& {Forestini}}]{Siess00}
{Siess}, L., {Dufour}, E., \& {Forestini}, M. 2000, \aap, 358, 593

\bibitem[{{Slesnick} {et~al.}(2006){Slesnick}, {Carpenter}, \&
  {Hillenbrand}}]{Slesnick06a}
{Slesnick}, C.~L., {Carpenter}, J.~M., \& {Hillenbrand}, L.~A. 2006, \aj, 131,
  3016

\bibitem[{{Stauffer} {et~al.}(1999){Stauffer}, {Barrado y Navascu{\'e}s},
  {Bouvier}, {Morrison}, {Harding}, {Luhman}, {Stanke}, {McCaughrean},
  {Terndrup}, {Allen}, \& {Assouad}}]{Stauffer99}
{Stauffer}, J.~R., {Barrado y Navascu{\'e}s}, D., {Bouvier}, J., {et~al.} 1999,
  \apj, 527, 219

\bibitem[{{Stauffer} {et~al.}(1998){Stauffer}, {Schultz}, \&
  {Kirkpatrick}}]{Stauffer98}
{Stauffer}, J.~R., {Schultz}, G., \& {Kirkpatrick}, J.~D. 1998, \apjl, 499,
  L199+

\bibitem[{{Vacca} {et~al.}(2003){Vacca}, {Cushing}, \& {Rayner}}]{Vacca03}
{Vacca}, W.~D., {Cushing}, M.~C., \& {Rayner}, J.~T. 2003, \pasp, 115, 389

\bibitem[{{Ventura} {et~al.}(1998){Ventura}, {Zeppieri}, {Mazzitelli}, \&
  {D'Antona}}]{Ventura98}
{Ventura}, P., {Zeppieri}, A., {Mazzitelli}, I., \& {D'Antona}, F. 1998, \aap,
  331, 1011

\bibitem[{{Whitworth} \& {Zinnecker}(2004)}]{Whitworth04}
{Whitworth}, A.~P. \& {Zinnecker}, H. 2004, \aap, 427, 299

\bibitem[{{Wilking} {et~al.}(1999){Wilking}, {Greene}, \& {Meyer}}]{Wilking99}
{Wilking}, B.~A., {Greene}, T.~P., \& {Meyer}, M.~R. 1999, \aj, 117, 469

\bibitem[{{Zapatero Osorio} {et~al.}(2002){Zapatero Osorio}, {B{\'e}jar},
  {Pavlenko}, {Rebolo}, {Allende Prieto}, {Mart{\'{\i}}n}, \& {Garc{\'{\i}}a
  L{\'o}pez}}]{Zapatero02}
{Zapatero Osorio}, M.~R., {B{\'e}jar}, V.~J.~S., {Pavlenko}, Y., {et~al.} 2002,
  \aap, 384, 937

\end{thebibliography}

\onllongtab{6}{
\begin{landscape}
\tiny
\begin{longtable}{@{\extracolsep{-10pt}}lllrccccccccccc}
\caption[Summary of the parameters derived in this work for Collinder 69 candidate members and comparison with previous studies.]{Summary of the parameters derived in this work for Collinder 69 candidate members and comparison with previous studies.\label{tab:paramTOTAL}}\\
\hline\hline
Object & Alt ID & RA & DEC & Membership$^1$ & SpT & T$_{\rm eff}$ & L$_{\rm bol}$(L$_{\odot}$) & M(M$_{\odot}$) & V$_{\rm Li}$ &  Ins$^4$ & EW(Li)$^5$ & EW(KI)   & EW(KI)   & EW(NaI)  \\
       &        &    &     &                &     &               &                            &                &         &          & 6708 \AA   & 7665 \AA & 7700 \AA & 8200 \AA  \\      
\hline
\endfirsthead

\multicolumn{5}{c}{{\tablename} \thetable{} -- \scriptsize Continued} \\
\hline\hline
Object         & Alt ID             & RA        & DEC       & Membership$^1$  & SpT         & T$_{\rm eff}$ & L$_{\rm bol}$(L$_{\odot}$) & M(M$_{\odot}$) & V$_{\rm Li}$ &  Ins$^4$ & EW(Li)$^5$                           & EW(KI)               & EW(KI)               & EW(NaI)  \\
               &                    &           &           &                 &             &               &                            &                &         &          & 6708 \AA                                  & 7665 \AA             & 7700 \AA             & 8200 \AA  \\      
\hline
\endhead

\hline \multicolumn{5}{r}{{Continued on next page\ldots}} \\ \hline
\endfoot

\hline\hline
\endlastfoot
  LOri001      & C69XW-004,  DM001  & 83.44658 & 9.92736 & Y B10           & M0                   & 4000 & 0.928$\pm$0.004   & 0.954 &   & 8        & 0.510 $\pm$ 0.007$^{1)}$ , 0.56$^{2)}$                            &                      &                      &                     \\
  LOri003      & C69XE-001,  DM046  & 83.98100 & 9.94208 & Y S08           & K8.5$^2$             & 4000 & 0.823$\pm$0.004   & 0.919 &   &          &                            0.55$^{2)}$   , 0.520$\pm$0.032$^{3)}$ &                      &                      &                     \\
  LOri007      &                    & 83.62313 & 9.81631 & NM B10          & M1.5 , M1.0          & 4000 & 0.627$\pm$0.003   & 0.836 &   & 8, 6     &                                                                   & 0.47 $\pm$0.213      & 0.693$\pm$0.11       & 3.195$\pm$0.306     \\
  LOri008      & C69XE-004,  DM051  & 83.99154 & 9.90911 & Y DM            &                      & 4000 & 0.719$\pm$0.004   & 0.877 &   &          &                            0.55$^{2)}$                            &                      &                      &                     \\
  LOri011      &                    & 83.68608 & 9.89931 & NM B10          & K9                   & 3750 & 0.673$\pm$0.003   & 0.737 &   & 8        & 0.038 $\pm$ 0.002$^{1)}$                                          &                      &                      &                     \\
  LOri012      &                    & 83.77479 & 9.86883 & NM B10          & K8                   & 4000 & 0.602$\pm$0.003   & 0.824 &   & 8        & 0.082 $\pm$ 0.008$^{1)}$                                          &                      &                      &                     \\
\end{longtable}

{\footnotesize
{\bf Notes:}\\
$^1$ Membership confirmed by:\\
DM - \citet{DM99,DM01} \\
S08 - \citet{Sacco08}  \\
M08 - \citet{Maxted08} \\
B10 - This work\\
$^2$Spectral type photometrically determined by \citet{Sacco08} (all the remainig spectral types were obtained spectroscopically in this work).\\
$^3$Infrared spectral type (see Section~\ref{IR_SpT}).\\
$^4$ Instrumental configuration as follows:\\
 (1) LRIS R$\sim$2650\\ 
 (2) LRIS R$\sim$950\\ 
 (3) MIKE R$\sim$11250\\
 (4) B\&C R$\sim$2600\\
 (5) B\&C R$\sim$800\\
 (6) TWIN R$\sim$1100\\
 (7) CAFOS R$\sim$600\\
 (8) FLAMES R$\sim$8600\\
 (9) NIRSPEC R$\sim$2000\\
 (10) SOFI R$\sim$950\\
 (11) IRCS R$\sim$200\\
$^5$ Measurement from:\\
$^{1)}$ This work \\
$^{2)}$ \citet{DM99,DM01}\\
$^{3)}$ \citet{Sacco08}  \\
}
\normalsize
\end{landscape}

}
\Online

\begin{appendix} 
\section{Automatic line characterization procedure and resolution effects.}

In short; one of the most difficult steps to automate when measuring line properties is the determination of the continuum on top of (or below) which the line lies. Even when this task is performed manually, different astronomers might select different points as the bases of the continuum when, for example, facing a noisy spectrum or a line embedded in a molecular band. The code that we have developed tries to solve this problem proceeding in an iterative and consistent way. 

\begin{figure}[!h]
\centering
\includegraphics[width=9cm]{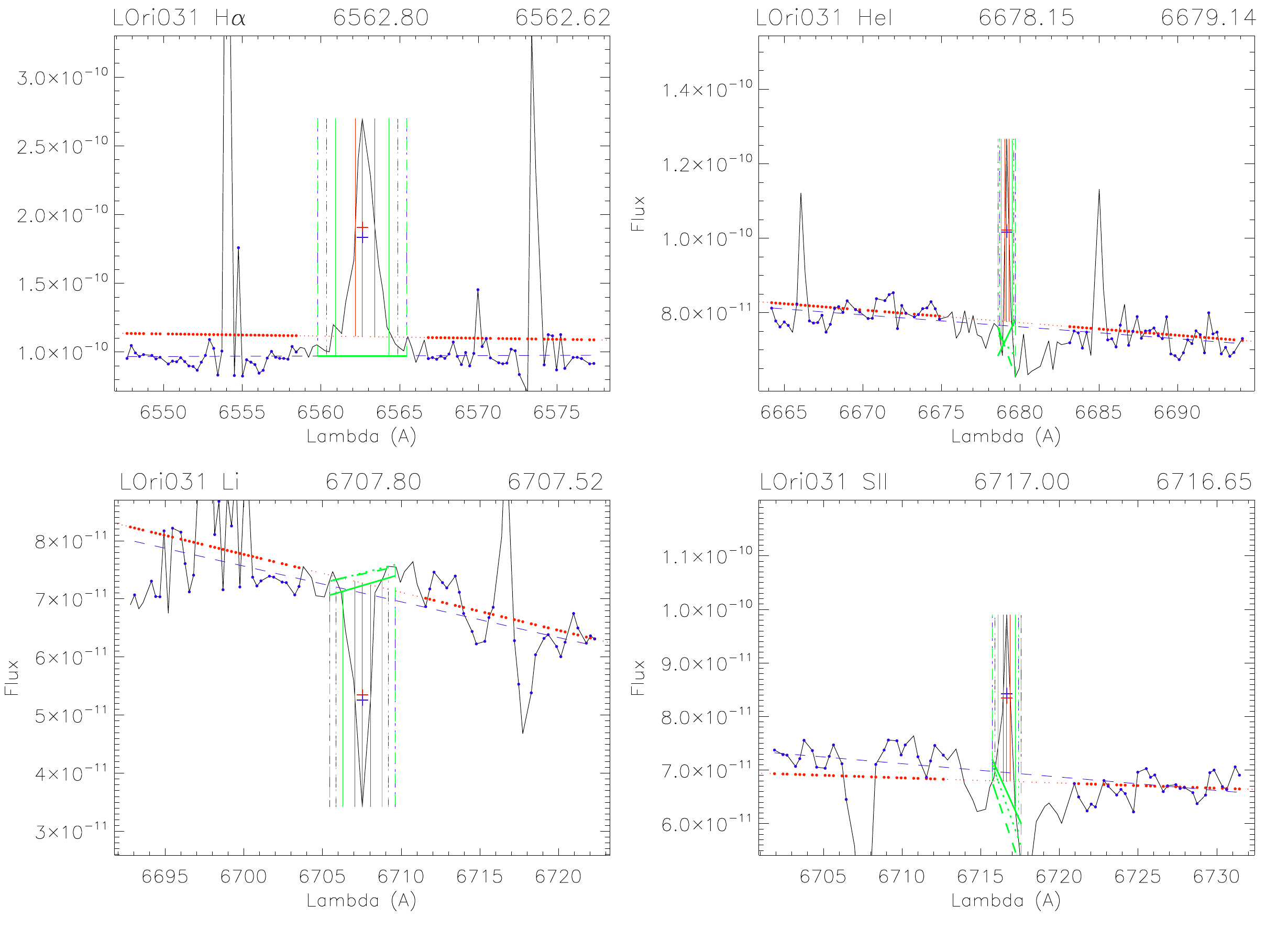}
\caption{Example of the output of the code designed to measure EW, FWHM and FW$_{10\%}$ of the different lines detected on the spectra in an automatic manner. In this case we show four of the measured lines on the averaged MIKE spectra obtained for LOri031: H$\alpha$ at 6562.80\AA, He I at 6678.15\AA, Li I at 6707.8\AA and S II at 6717.0\AA. For each case the canonical wavelength for the measured line and the actual wavelength where the line has been found in our spectra are displayed o the title of the plot (all of them within the errors on wavelength calibration). The iterative process in the continiuum determination is displayed using different colours: red for the first iteration (including the first guess for the half maximum flux value, red cross), blue for the second one (filled blue dots on the data-points used to refine the first continiuum, blue cross showing the second half maximum estimation) and the three green lines (filled, dotted and dashed) representing the final determined continuum.}
\label{ex_lines}
\end{figure}

First of all it locates where the actual peak of the line is; in this step the wavelength calibration accuracy is taken into account providing limits in the wavelength coverage where the search for the peak of the line is carried out. Once this task is accomplished, two regions are defined (to the right and to the left of the estimated central wavelength) the width of the regions is fixed to 100 points to be statistically significant (and therefore the width in wavelength units would depend on the resolution of the spectra). A linear fit using those regions is performed (red dotted line in Fig~\ref{ex_lines}). With this attempt to define a continiuum (the linear fit), the code calculates a first guess of the FWHM (red cross in Fig~\ref{ex_lines}). 

For the second iteration, the code considers that the line is a gaussian, and, therefore, $\rm{FWHM} = 2 \times \sqrt{2 \times \log(2)} \times \sigma$. Assuming that at a distance of $\pm 10\sigma$ one should be outside the line, two new regions on the spectra are selected: starting at $\pm 10\sigma$ from the wavelength of the peak and ending at the same limits as before. These two regions are close to the edges of the line unless the line has a very wide double peaked structure.

At this point the code will perform two linear fits in a sequential way; a second continuum is derived with a linear fit to these new regions of the spectra close to the edges of the line, and the third one with another lineal fitting, but this time considering only those data-points that differ from the second continuum less than once the dispersion of the difference between the actual spectra and this second continuum (the blue dots in Fig~\ref{ex_lines} represent those data-points considered to define a third continuum and the blue dashed line is the resulting one). 

Once the third guess on the locus of the continuum is estimated, the code calculates a second iteration of the FWHM (blue cross in Fig~\ref{ex_lines}). Assuming one more time that the line can be described with a gaussian (and calculating the associated $\sigma$), it considers three pairs of lambdas in the spectra to make the final measurements (pairs located at $\pm 3\sigma$, $\pm 4\sigma$ and $\pm 5\sigma$ from the lambda of the peak). As a final refinement, for each selected pair of wavelengths; for example $\lambda_{1}$ and $\lambda_{2}$ corresponding to $\lambda_{\rm peak}\pm 3\sigma$ the code checks whether the linear continuum defined by ($\lambda_{1}$,F($\lambda_{1}$)), ($\lambda_{2}$,F($\lambda_{2}$)) intersects the wings of the line; when this is the case (for example the upper-left panel of Fig~\ref{ex_lines}), the code will use the third continuum to perform the measurements. 

Prior to provide these measurements, the code subtracts the instrumental profile from the values estimated for the FWHM and FW$_{10\%}$ considering that the FWHM of the convolution (G$_{\rm measured}$) of two gaussians (G$_{\rm instr}$, G$_{\rm line}$) follows: 
\begin{equation}
{\rm FWHM(G}_{\rm measured}) =  \sqrt{{\rm FWHM(G}_{\rm instr})^{2}+{\rm FWHM(G}_{\rm line})^{2}}
\end{equation}
and that FWHM and FW$_{10\%}$ relate accordig to: FW$_{10\%}$ = $\sqrt{\frac{\log(2)}{\log(10)}}$ $\times$ FWHM.

The final product is the mean of the three measurements for each parameter (FWHM, FW$_{10\%}$ and EW) and their corresponding standard deviation $\sigma$. 

\begin{center}
\begin{figure*}
\includegraphics[width=17.0cm]{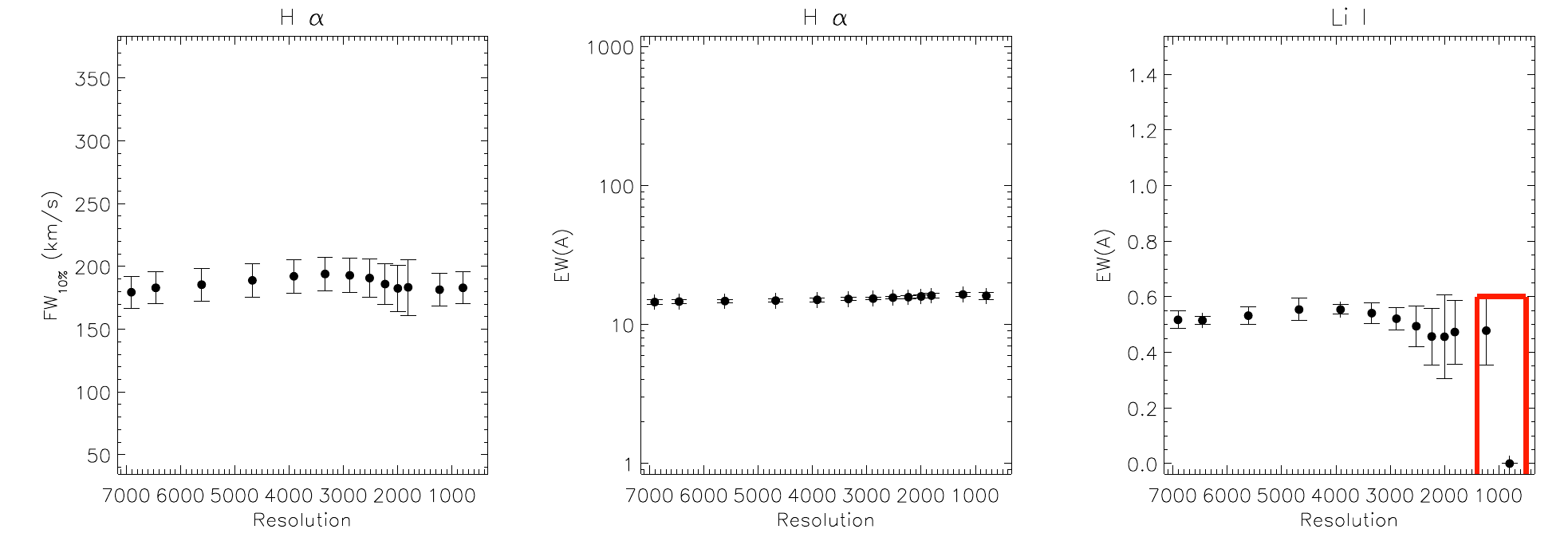}
\caption{Relationship between different measurements of the H$\alpha$ and Li I line profiles and the resolution os the spectra where the measurement has been performed.}
\label{fig:EW_res}
\end{figure*}
\end{center}

\begin{center}
\begin{figure*}
\includegraphics[width=18.0cm]{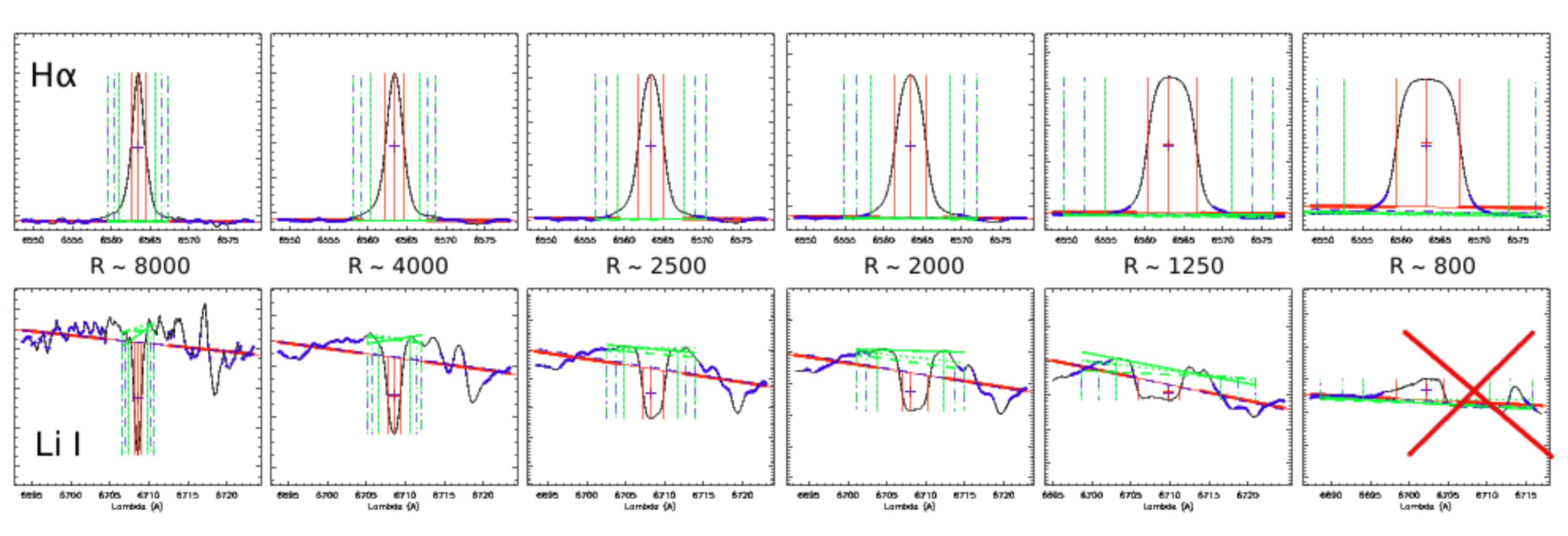}
\caption{Sequence composed by a detailed view of the evolution of the profiles of the H$\alpha$ (emission) and Li I (absorption) lines with the variation of the spectral resolution. The first panel (from left to right) corresponds to the original resolution of our FLAMES spectrum of LOri038, a M3 member of C69 (R $\sim$8000). In the last panel we have degraded the FLAMES spectra to a resolution similar to that of our low-resolution campaign (R $\sim$800 obtained with the B\&C spectrograph). As can be seen in this figure and in Fig.~\ref{fig:EW_res} the lithium's equivalent width can be measured down to resolutions of the order of 1250 (for a good signal to noise ratio as it is the case of our spectrum of LOri038). The linestyles and color code are those explained on the text.}
\label{fig:deg_res}
\end{figure*}
\end{center}

Once we had an automatic procedure to characterize the lines present in our spectra, we had to address the effect that the wide dispersion in resolution might introduce in our measurements. We proceeded in the following manner: We degraded one of our FLAMES spectra (from LOri038, a M3, accreting, Class II, confirmed member) to different resolutions (including those listed in Table~\ref{C69campains}). We used then (on those degraded spectra) our code to measure FHWM, EW and FW$_{10\%}$ for the H$\alpha$ (emission) and Li $\lambda$6708~\AA~ lines (absorption). In Fig.~\ref{fig:EW_res} we provide those values (EW and FW$_{10\%}$ for H$\alpha$ and EW for Li I $\lambda$6708~\AA) as a function of the resolution of the spectra. The length of the y-axes has been fixed by purpose for each plot: in the first panel (from left to right), it represents the variation of the instrumental response (the FHWM measured on the respective arc adapted for each value of the resolution) among the resolutions considered. In the cases of the middle and right-side panels, the length of this axes provides an idea of the variation within members of Collinder 69 for the corresponding equivalent width. Note that in the case of H$\alpha$ there is clearly no significant dependence on the measurements made at different resolutions (the variations lie within the error bars). For the case of the lithium, due to the intrinsic weakness of the line, the accuracy of the measurement (meaning the error-bars) lowers with the spectral resolution, but still it seems perfectly reasonable to compare measurements at R $\sim$ 2000 and R $\sim$ 8000. In the same figure we have highlighted the range of resolutions where Li is not detectable anymore. A detailed view of the ``degeneration'' of the lines for this example is provided in fig.~\ref{fig:deg_res}

\end{appendix}

\end{document}